\documentclass[12pt]{article}
\begin{document}

\setcounter{page}{1}
\def\theequation{\arabic{section}.\arabic{equation}}
\def\theequation{\thesection.\arabic{equation}}
\setcounter{section}{0}

\title{The Nambu--Jona--Lasinio model of light nuclei}

\author{A.N. Ivanov\thanks{E--mail: ivanov@kph.tuwien.ac.at, Tel.:
+43--1--58801--14261, Fax: +43--1--58801--14299}~$^{\ddagger}$,
H. Oberhummer\thanks{E--mail: ohu@kph.tuwien.ac.at, Tel.:
+43--1--58801--14251, Fax: +43--1--58801--14299} ,
N.I. Troitskaya\thanks{Permanent address: State Technical
University, Department of Nuclear Physics, 195251 St.Petersburg} ,\\
M. Faber\thanks{E--mail: faber@kph.tuwien.ac.at, Tel.:
+43--1--58801--14261, Fax: +43--1--58801--14299}}

\date{\today}
\maketitle
\vspace{-0.5in}

\begin{center}
{\it Institut f\"ur Kernphysik, Technische Universit\"at Wien,\\
Wiedner Hauptstr. 8--10/142, A--1040, Vienna, Austria}
\end{center}

\begin{abstract}
The Nambu--Jona--Lasinio model of the deuteron suggested by Nambu and
Jona--Lasinio (Phys. Rev. 124 (1961) 246) is formulated from the first
principles of QCD. The deuteron appears as a neutron--proton
collective excitation, i.e. a Cooper np--pair, induced by a
phenomenological local four--nucleon interaction in the nuclear phase
of QCD. The model describes the deuteron coupled to itself, nucleons
and other particles through one--nucleon loop exchanges providing a
minimal transfer of nucleon flavours from initial to final nuclear
states and accounting for contributions of nucleon--loop anomalies
which are completely determined by one--nucleon loop diagrams. The
dominance of contributions of nucleon--loop anomalies to effective
Lagrangians of low--energy nuclear interactions is justified in the
large $N_C$ expansion, where $N_C$ is the number of quark colours.
\end{abstract}

\begin{center}
\end{center}
\newpage

\section{Introduction}
\setcounter{equation}{0}

\hspace{0.2in} In the beginning of sixties Nambu and Jona--Lasinio
suggested a dynamical theory of elementary particles [1,2], the
Nambu--Jona--Lasinio (NJL) model, in which nucleons and mesons are
derived in a unified way from a fundamental spinor field on the basis
of the relativistic extension of the BCS (Bardeen--Cooper--Schrieffer)
theory of superconductivity [3]. Nowadays the NJL model has found a
great support in the form of the extended Nambu--Jona--Lasinio (ENJL)
model with chiral $U(3)\times U(3)$ symmetry as an effective
phenomenological approximation of low--energy Quantum Chromodynamics
(QCD) [4--6]. Chiral perturbation theory within the ENJL model with a
linear and a non--linear realization of chiral $U(3)\times U(3)$
symmetry has been developed in Refs.\,[7,8] and [9], respectively. In
the ENJL model mesons are described as $q\bar{q}$ collective
excitations, the $q\bar{q}$ Cooper pairs, induced by phenomenological
local four--quark interactions. In turn, the low--lying octet and
decuplet of baryons can be considered in the NJL approach as
three--quark $qqq$ collective excitations produced by phenomenological
local six--quark interactions [8]. As has been shown in Refs.\,[4--11]
the ENJL model with chiral $U(3)\times U(3)$ symmetry describes at the
quark level perfectly well strong low--energy interactions of hadrons
in the form of Effective Chiral Lagrangians [12,13].

In parallel to the description of mesons as a collective excitations
of a unified spinor field Nambu and Jona--Lasinio suggested to treat
the deuteron as a neutron--proton collective excitation, i.e. some
kind of a Cooper np--pair [2]. A phenomenological local four--nucleon
interaction has been written in the form [2]:
\begin{eqnarray}\label{label1.1}
{\cal L}_{\rm int}(x)
&=&\frac{1}{4}\,g_0\,[\bar{\psi}(x)\gamma_{\mu}\psi^c(x)\bar{\psi^c}(x)
\gamma^{\mu}\psi(x) +
\bar{\psi}(x)\sigma_{\mu\nu}\psi^c(x)\bar{\psi^c}(x)
\sigma^{\mu\nu}\psi(x)\nonumber\\
&&\hspace{0.2in} + 
\bar{\psi}(x)\gamma_{\mu}\gamma^5\vec{\tau}
\psi^c(x)\cdot\bar{\psi^c}(x) \gamma^{\mu}\gamma^5\vec{\tau}\psi(x)],
\end{eqnarray}
where $\psi(x)$ is a doublet of the nucleon field, then $\psi^c(x) =
C\,\bar{\psi}^T(x)$ and $\bar{\psi^c}(x) = \psi^T(x)\,C$, $C$ is a
charge conjugation matrix and $T$ is a transposition; $\vec{\tau} =
(\tau^1, \tau^2, \tau^3)$ are the isotopical Pauli matrices, and
$\sigma_{\mu\nu} = (\gamma_{\mu}\gamma_{\nu} -
\gamma_{\nu}\gamma_{\mu})/2$.

According to the Nambu--Jona--Lasinio prescription [2] the
phenomenological interactions Eq.(\ref{label1.1}) should lead to two
bound states: a pseudovector, isoscalar ($J = 1^+, I = 0$) or the
deuteron, where $J$ and $I$ are the total spin and isospin,
respectively, coming from the first two interaction terms, and a
scalar, isovector ($J = 0^+, I = 1$) coming from the last term.
Unfortunately, Nambu and Jona--Lasinio did not consider the evaluation
of the binding energy, the magnetic dipole and electric quadrupole
moments of the deuteron in their approach to the deuteron as the
Copper np--pair. Such attempts have been undertaken within the model
which has been called the Relativistic field theory model of the
deuteron (RFMD) suggested in Refs.\,[14,15]. The RFMD realizes the
consideration of the deuteron in the analogous way to the
Nambu--Jona--Lasinio approach [2].  For the practical evaluation of
the low--energy parameters characterizing the deuteron there has been
suggested in the RFMD that the deuteron couples to itself, nucleons
and other particles through one--nucleon loop exchanges providing a
minimal transfer of nucleon flavours from initial to final nuclear
states and accounting for contributions of nucleon--loop anomalies
which are completely determined by one--nucleon loop diagrams. Then,
there has been argued the dominant role of one--nucleon loop anomalies
for the one--nucleon loop exchanges describing strong low--energy
nuclear forces. The main problem of the attempts expounded in
Refs.\,[14,15] as well as the original idea of the Nambu and
Jona--Lasinio to describing the deuteron as the Cooper np--pair [2] is
in a poor relation to QCD.

In this paper we show that the consideration of the deuteron as a
Cooper np--pair induced by a phenomenological local four--nucleon
interaction and the description of low--energy couplings of the
deuteron to itself, nucleons and other particles through one--nucleon
loop exchanges, where nucleon--loop anomalies play the dominant role,
is fully motivated by low--energy QCD. The deuteron appears as a
Cooper np--pair in the nuclear phase of QCD and couples to itself and
other particles through the one--nucleon loop exchanges. The dominance
of nucleon--loop anomalies occurs naturally as a consequence of the
large $N_C$ expansion, where $N_C$ is the number of quark colour
degrees of freedom [16,17]. Nowadays the large $N_C$ expansion
suggested by 't Hooft [16] is accepted as a non--perturbative approach
of low--energy QCD to the analysis of strong couplings of hadrons and
nuclei at low energies [17].

Below we would call the Nambu--Jona--Lasinio approach to the deuteron
based on phenomenological local four--nucleon interactions like those
given by Eq.(\ref{label1.1}) as the Nambu--Jona--Lasinio model of
light nuclei or shortly the nuclear Nambu--Jona--Lasinio model with
the abbreviation the NNJL model.

The paper is organized as follows. In Section 2 we discuss the
non--perturbative phases of QCD and formulate the NNJL model from the
first principles of QCD. In Section 3 we derive the effective
Lagrangian for the free deuteron field induced in the nuclear phase of
QCD as the neutron--proton collective excitation (the Cooper np--pair)
by a phenomenological local four--nucleon interaction.  We demonstrate
the dominant role of one--nucleon loop anomalies for the formation of
the effective Lagrangian of the free deuteron field. In Section 4 we
investigate the electromagnetic properties of the deuteron and derive
the effective Lagrangian of the deuteron field coupled to an external
electromagnetic field through the magnetic dipole and electric
quadrupole moments. We show that the effective Lagrangian of the
electromagnetic interactions of the deuteron calculated in the
one--nucleon loop approximation at leading order in the large $N_C$
expansion are defined by the anomalies of one--nucleon loop diagrams
and have the form of well--known phenomenological electromagnetic
interactions introduced by Corben and Schwinger [18] and Aronson [19]
for the description of charged vector boson fields coupled to an
external electromagnetic field. In the Conclusion we discuss the
obtained results.

\section{Non--perturbative phases of QCD}
\setcounter{equation}{0}

\hspace{0.2in} The derivation of the NNJL model from the first principles of
QCD goes through three non--perturbative phases of the
quark--gluon system. We call them as: 1) the low--energy quark--gluon
phase (low--energy QCD), 2) the hadronic phase and 3) the nuclear
phase.

{\it The low--energy quark--gluon phase of QCD} can be obtained by
integrating over fluctuations of quark and gluon fields at energies
above the scale of spontaneous breaking of chiral symmetry (SB$\chi$S)
$\Lambda_{\chi} \simeq 1\,{\rm GeV}$.  This results in an effective
field theory, low--energy QCD, describing strong low--energy
interactions of quarks and gluons. The low--energy quark--gluon phase
of QCD characterizes itself by the appearance of low--energy
gluon--field configurations leading to electric colour fluxes
responsible for formation of a linearly rising interquark
potential. The former realizes quark confinement.

For the transition to {\it the hadronic phase of QCD} one should,
first, integrate out low--energy gluon degrees of freedom.
Integrating over gluon degrees of freedom fluctuating around
low--energy gluon--field configurations responsible for formation of a
linearly rising interquark potential one arrives at an effective field
theory containing only quark ($q$) and anti--quark ($\bar{q}$) degrees
of freedom. This effective field theory describes strong interactions
at energies below the SB$\chi$S scale $\Lambda_{\chi} \simeq 1\,{\rm
GeV}$.  The resultant quark system possesses both a chirally invariant
and chirally broken phase. In the chirally invariant phase the
effective Lagrangian of the quark system is invariant under chiral
$U(3)\times U(3)$ group. The chirally invariant phase of the quark
system is unstable and the transition to the chirally broken phase is
advantageous. The chirally broken phase characterizes itself by three
non--perturbative phenomena: SB$\chi$S, hadronization (creation of
bound quark states with quantum numbers of mesons $q\bar{q}$,
$qq\bar{q}\bar{q}$, baryons $qqq$ and so on) and confinement. The
transition to the chirally broken phase caused by SB$\chi$S
accompanies itself with hadronization. Due to quark confinement all
observed bound quark states should be colourless. As gluon degrees of
freedom are integrated out, in such an effective field theory the
entire variety of strong low--energy interactions of hadrons at
energies below the SB$\chi$S scale $\Lambda_{\chi}\simeq 1\,{\rm GeV}$
is described by quark--loop exchanges.

Since nowadays in continuum space--time formulation of QCD the
integration over low--energy gluon--field configurations can be hardly
performed explicitly, phenomenological approximations of this
integration represented by different effective quark models with
chiral $U(3)\times U(3)$ symmetry motivated by QCD are welcomed.

The most interesting effective quark model allowing to describe
analytically both SB$\chi$S and bosonization (creation of bound
$q\bar{q}$ states with quantum numbers of observed low--lying mesons)
is the extended Nambu--Jona--Lasinio (ENJL) model [4--11] with linear
[7,8,10] and non--linear [9,11] realization of chiral $U(3)\times
U(3)$ symmetry. As has been shown in Ref.\,[6] the ENJL model is fully
motivated by low--energy multi--colour QCD with a linearly rising
interquark potential and $N_C$ quark colour degrees of freedom at $N_C
\to \infty$. In the ENJL model mesons are described as $q\bar{q}$
collective excitations (the Cooper $q\bar{q}$--pairs) induced by
phenomenological local four--quark interactions. Through
one--constituent quark--loop exchanges the Copper $q\bar{q}$--pairs
acquire the properties of the observed low--lying mesons such as
$\pi(140)$, $K(498)$, $\eta(550)$, $\rho(770)$, $\omega(780)$,
$K^*(890)$ and so on. For the description of low--lying octet and
decuplet of baryons the ENJL model has been extended by the inclusion
of local six--quark interactions responsible for creation of baryons
as $qqq$ collective excitations [8].

Integrating then out low--energy quark--field fluctuations, that can
be performed in terms of constituent quark--loop exchanges, one
arrives at {\it the hadronic phase of QCD} containing only local meson
and baryon fields. The couplings of low--lying mesons and baryons are
described by Effective Chiral Lagrangians with chiral $U(3)\times
U(3)$ symmetry [4--13].

{\it The nuclear phase of QCD} characterizes itself by the appearance
of bound nucleon states -- nuclei. In order to arrive at {\it the
nuclear phase of QCD} we suggest to start with the hadronic phase of
QCD and integrate out heavy hadron degrees of freedom, i.e. all heavy
baryon degrees of freedom with masses heavier than masses of
low--lying octet and decuplet of baryons and heavy meson degrees of
freedom with masses heavier than the SB$\chi$S scale $\Lambda_{\chi}
\simeq 1\,{\rm GeV}$.  At low energies the result of the integration
over these heavy hadron degrees of freedom can be represented in the
form of phenomenological local many--nucleon interactions. Following
the scenario of the hadronic phase of QCD, where hadrons are produced
by phenomenological local many--quark interactions as many--quark
collective excitations, one can assume that some of these
many--nucleon interactions are responsible for creation of
many--nucleon collective excitations. These excitations acquire the
properties of observed bound nucleon states -- nuclei through
nucleon--loop and low--lying meson exchanges. This results in an
effective field theory describing nuclei and their low--energy
interactions in analogy with Effective Chiral Lagrangian approaches
[12,13].  Chiral perturbation theory  can be  naturally
incorporated into this effective field theory of low--energy
interactions of nuclei.

We would like to emphasize that in this scenario of the quantum field
theoretic formation of nuclei and their low--energy interactions
nuclei are considered as elementary particles described by local
interpolating fields.  In parallel to the Nambu--Jona--Lasinio
approach to light nuclei [2] the representation of nuclei as
elementary particles has been suggested by Sakita and Goebal [20] and
Kim and Primakoff [21] for the description of electromagnetic and weak
nuclear processes. We develop the quantum field theoretic approach to
the interpretation of nuclei as elementary particles represented by
local interpolating fields by starting with QCD.

In this scenario the deuteron, being the lightest bound nucleon state,
appears in the nuclear phase of QCD as the neutron--proton collective
excitation (the Cooper np--pair) induced by a phenomenological local
four--nucleon interaction caused by the contributions of heavy hadron
exchanges at low energies. Through one--nucleon loop exchanges the
Cooper np--pair with quantum numbers of the physical deuteron acquires
the properties of the physical deuteron (i) the binding energy
$\varepsilon_{\rm D} = 2.225\,{\rm MeV}$, (ii) the magnetic dipole
moment $\mu_{\rm D} = 0.857\,{\mu}_{\rm N}$, where ${\mu}_{\rm N}$ is
a nuclear magneton, (iii) the electric quadrupole moment $Q_{\rm
D} = 0.286\,{\rm fm}^2$ [22] and so on.

We would like to emphasize that Sakita and Goebal [20] by treating the
deuteron as an elementary particle described by a local interpolating
field $D_{\mu}(x)$ have calculated the cross section for the
photo--disintegration of the deuteron $\gamma$ + D $\to$ n + p within
the dispersion relation approach. The more recent analysis of the same
process by using the dispersion relations has been carried out by
Anisovich and Sadovnikova [23] based on the dispersion relation
technique developed by Anisovich {\it et al.} [24]. The dispersion
relation approach as well as the NNJL model is a relativistically
covariant one. Within the dispersion relation approach one deals with
directly the amplitudes of the process of the deuteron coupled to
other particles keeping under the control intermediate states in the
form of the pole and branching point singularities. The residues at
pole singularities are defined by the effective coupling constants
which as usual in the dispersion relation technique are taken from
experimental data. Unlike the dispersion relation approach in the NNJL
model developing the Lagrange approach to nuclear forces we focus on
the evaluation of the effective coupling constants via the derivation
of the effective Lagrangians of the deuteron coupled to nucleons and
other particles at low energies. These effective coupling constants
are defined in the NNJL model by one--nucleon loop anomalies related
to high--energy $N\bar{N}$ fluctuations of virtual nucleon $(N)$ and
anti--nucleon $(\bar{N})$ fields. Thus, the dispersion relation
approach to the description of low--energy interactions of the
deuteron and the NNJL model do not contradict but should complement
each other.

In the form of the path integral formulation of QCD the
non--perturbative phases of QCD can be represented by the following
sequence of transformations. Let us start with the path integral over
the quark $q$, anti--quark $\bar{q}$ and gluon $A$ fields related to a
generating functional of quark and gluon Green functions and defined
by
\begin{eqnarray}\label{label2.1}
{\cal Z} = \int Dq D\bar{q}DA\, e^{\textstyle i\int
d^4x\,{\cal L}^{\rm QCD}[\bar{q}, q, A]}.
\end{eqnarray}
Integrating over high--energy quark--gluon fluctuations restricted
from below by the SB$\chi$S scale $\Lambda_{\chi} \simeq 1\,{\rm GeV}$
we arrive at the generating functional
\begin{eqnarray}\label{label2.2}
{\cal Z} = \int Dq
D\bar{q}Da\,e^{\textstyle i\int d^4x\,{\cal L}^{\rm QCD}_{\rm
eff}[\bar{q}, q,\tilde{A} + a]}
\end{eqnarray}
describing strong low--energy interactions of quarks and gluons in
{\it the low--energy quark--gluon phase of QCD}, where $\tilde{A}$ and
$a$ are non--perturbative gluon--field configurations responsible for
the formation of a linearly rising interquark potential providing
quark confinement and the gluon--field fluctuations around these
gluon--field configurations.

Integrating then over the gluon--field fluctuations $a$ we obtain the
generating functional
\begin{eqnarray}\label{label2.3}  
{\cal Z} = \int Dq
D\bar{q}\,e^{\textstyle i\int d^4x\,{\cal L}_{\rm eff}[\bar{q},
q, {local~multi-q~couplings}]}.
\end{eqnarray}
This generating functional describes the effective theory of quarks
coupled to each other at energies of order $\Lambda_{\chi} \simeq
1\,GeV$ and less. At the phenomenological level the result of the
integration over gluon--field configurations can be represented in the
form of phenomenological local multi--quark interactions responsible
for the creation of multi--quark collective excitations. This quark
system is unstable under SB$\chi$S and hadronization. By converting
quark degrees of freedom into the hadronic ones or differently
hadronizing the quark system we arrive at the generating functional
given by the path integral over hadronic degrees of freedom only
\begin{eqnarray}\label{label2.4}
{\cal Z} = \int
DM_{\ell} DB_{\ell} DM_h DB_h \,e^{\textstyle i\int d^4x\,{\cal
L}_{\rm eff}[M_{\ell},B_{\ell}, M_h, B_h]},
\end{eqnarray}
where $M_{\ell}$, $B_{\ell}$ and $M_h$, $B_h$ are local interpolating
fields of mesons and baryons. The indices ${\ell}$ and $h$ correspond
to light hadrons with masses of order of 1\,GeV and less and heavy
hadrons with masses much greater than 1\,GeV. For the practical
applications to the description of low--energy couplings of light and
heavy hadrons the effective Lagrangian ${\cal L}_{\rm
eff}[M_{\ell},B_{\ell}, M_h, B_h]$ can be approximated by Effective
Chiral Lagrangians for light hadrons [12,13] and heavy hadrons [25] as
well. The generating functional Eq.(\ref{label2.4}) describes
low--energy interactions of hadrons in  {\it the hadronic phase of QCD}.

Integrating over heavy baryon degrees of freedom given by the fields
$M_h$ and $B_h$ we get the generating functional in the form of the
path integral over the light hadron degrees of freedom
\begin{eqnarray}\label{label2.5}
{\cal Z} = \int DM_{\ell} DB_{\ell}
\,e^{\textstyle i\int d^4x\,{\cal L}_{\rm eff}[M_{\ell},B_{\ell}, {
local~multi-B_{\ell}~couplings}]}.
\end{eqnarray}
At low energies the result of the integration over heavy hadron
degrees of freedom can be represented phenomenologically by local
multi--baryon couplings some of which should be responsible for the
creation of multi--baryon excitations with quantum numbers of
nuclei. In term of the local interpolating fields of nuclei the
generating functional Eq.(\ref{label2.5}) acquires the form
\begin{eqnarray}\label{label2.6}
\hspace{-0.5in}{\cal Z} &=& \int D M_{\ell}¸\, D B_{\ell}\, D\,{\rm D}\, D\, {^3}{\rm
H}\, D\, {^3}{\rm He}\, D\, {^4}{\rm He}\ldots \,\nonumber\\
\hspace{-0.5in}&&\times\,e^{\textstyle i\int
d^4x\,{\cal L}_{\rm eff}[M_{\ell}, B_{\ell}, {\rm D}, {^3}{\rm H},
{^3}{\rm He}, {^4}{\rm He}, \ldots]},
\end{eqnarray}
where D, ${^3}{\rm H}$, ${^3}{\rm He}$ and ${^4}{\rm He}$ are the
local interpolating fields of the deuteron, the triton, the helium--3
and the helium--4, respectively. The ellipses stand for a possible
contribution of other nuclei.  The generating functional
Eq.(\ref{label2.6}) describes {\it the nuclear phase of QCD}, when
nuclei couple to each other and light hadrons at low energies. Chiral
perturbation theory [7,8] is naturally incorporated into this theory.

\section{The deuteron as a Cooper np--pair}
\setcounter{equation}{0}

\hspace{0.2in} In order to describe the deuteron as a Cooper
np--pair we introduce a phenomenological local four--nucleon
interaction caused by the integration over heavy hadron degrees of
freedom. First, let us consider the simplest form of this local
four--nucleon interaction 
\begin{eqnarray}\label{label3.1}
{\cal L}_{\rm int}(x) = - \frac{g^2_{\rm V}}{4M^2_{\rm
N}}\,j^{\dagger}_{\mu}(x)j^{\mu}(x),
\end{eqnarray}
where $g_{\rm V}$ is the phenomenological coupling constant of the
NNJL model [14,15], $M_{\rm N} = 940\,{\rm MeV}$ is the nucleon
mass. We neglect here the electromagnetic mass difference for the
neutron and the proton.  As has been found in [14,15] the coupling
constant $g_{\rm V}$ is related to the electric quadrupole moment of
the deuteron $Q_{\rm D}$: $g^2_{\rm V}= 2\pi^2 Q_{\rm D}M^2_{\rm N}$
[15].

The nucleon current $j^{\mu}(x)$ with the quantum numbers of the
deuteron is defined by [14,15]
\begin{eqnarray}\label{label3.2}
j^{\mu}(x)= -i\,[\bar{p^c}(x)\gamma^{\mu}n(x) -
\bar{n^c}(x)\gamma^{\mu}p(x)].
\end{eqnarray}
Here $p(x)$ and $n(x)$ are the interpolating fields of the proton and
the neutron, $N^c(x) = C\,\bar{N}^T(x)$ and $\bar{N^c}(x) =
N^T(x)\,C$, where $C$ is a charge conjugation matrix and $T$ is a
transposition.  In terms of the electric quadrupole moment of the
deuteron the phenomenological local four--nucleon interaction
Eq.(\ref{label3.1}) reads
\begin{eqnarray}\label{label3.3}
{\cal L}_{\rm int}(x) = - \frac{1}{2}\,\pi^2\,Q_{\rm
D}\,j^{\dagger}_{\mu}(x)j^{\mu}(x).
\end{eqnarray}
Now let us discuss the behaviour of the phenomenological coupling
constant $g^2_{\rm V}/4M^2_{\rm N}$ from the point of view of the
large $N_C$ expansion in QCD with the $SU(N_C)$ gauge group at $N_C
\to \infty$ [16,17]. Suppose, for simplicity, that the
phenomenological four--nucleon interaction Eq.(\ref{label3.1}) is
caused by exchanges of the scalar $f_0(980)$ and $a_0(980)$ mesons
being the lightest states among heavy hadrons we have integrated out.

Through a linear realization of chiral $U(3)\times U(3)$ symmetry and
the Goldberger--Treiman relation one can find that the coupling
constant of $\sigma$--mesons $g_{\rm \sigma NN}$, the
$q\bar{q}$--scalar mesons, coupled to the octet of low--lying baryons
should be of order $g_{\rm \sigma NN} \sim O(\sqrt{N_C})$ at $N_C \to
\infty$. The scalar mesons $f_0(980)$ and $a_0(980)$ are most likely
four--quark states with $qq\bar{q}\bar{q}$ quark structure [26,27]. In
the limit $N_C \to \infty$ such $qq\bar{q}\bar{q}$ states are
suppressed by a factor $1/N_C$ [17]. Thus, an effective coupling
constant of low--energy NN interaction caused by the
$qq\bar{q}\bar{q}$ scalar meson exchanges should be of order
$O(1/N_C)$ at $N_C \to \infty$. By taking into account that in QCD
with $N_C \to \infty$ the nucleon mass $M_{\rm N}$ is proportional to
$N_C$ [17], $M_{\rm N} = N_C M_q $, where $M_q \sim 300\,{\rm MeV}$ is
the constituent quark mass, we can introduce the nucleon mass $M_{\rm
N}$ in the effective coupling constant as a dimensional parameter
absorbing the factor $N^2_C$, i.e. $g^2_{\rm V}/4M^2_{\rm N}$.  This
is also required by the correct dependence of the deuteron mass on
$N_C$.  As a result the phenomenological coupling constant $g_{\rm V}$
turns out to be of order $O(\sqrt{N_C})$ at $N_C \to \infty$.

We should emphasize that one does not need to know too much about
quark structure of heavy hadron degrees of freedom we have integrated
out. Without loss of generality one can argue that among the multitude
of contributions caused by the integration over heavy hadron degrees
of freedom one can always find the required local four--nucleon
interaction the effective coupling constant of which behaves like
$O(1/N_C)$ at $N_C \to \infty$. As we show below this behaviour of the
coupling constant of the phenomenological four--nucleon interaction
leads to the deuteron as bound neutron--proton state, and it is also
consistent with the large $N_C$ dependence of low--energy parameters
of the physical deuteron [17].

The effective Lagrangian of the np--system unstable under creation of
the Cooper np--pair with quantum numbers of the deuteron
is then defined by
\begin{eqnarray}\label{label3.4}
{\cal L}^{\rm np}(x)&=&\bar{n}(x)\,(i\gamma^{\mu}\partial_{\mu} - M_{\rm
N})\,n(x)+ \bar{p}(x)\,(i\gamma^{\mu}\partial_{\mu} -
M_{\rm N})\,p(x)  \nonumber\\ 
&&- \frac{g^2_{\rm V}}{4M^2_{\rm
N}}\,j^{\dagger}_{\mu}(x)j^{\mu}(x) + \ldots,
\end{eqnarray}
where ellipses stand for low--energy interactions of the neutron and the
proton with other fields.  

In order to introduce the interpolating local deuteron field we should
linearalize the Lagrangian Eq.(\ref{label3.4}). Following the
procedure described in Refs.\,[4--11] for the inclusion of local
interpolating meson fields in the ENJL model we get
\begin{eqnarray}\label{label3.5}
&&{\cal L}^{\rm np}(x) \to \bar{n}(x)\,(i\gamma^{\mu}\partial_{\mu} -
M_{\rm N})\,n(x) + \bar{p}(x)\,(i\gamma^{\mu}\partial_{\mu} - M_{\rm
N})\,p(x) \nonumber\\ 
&&+M^2_0\,D^{\dagger}_{\mu}(x)D^{\mu}(x) +
g_{\rm V}j^{\dagger}_{\mu}(x)D^{\mu}(x) + g_{\rm
V}j^{\mu}(x)D^{\dagger}_{\mu}(x) + \ldots,
\end{eqnarray}
where $M_0 = 2\,M_{\rm N}$ and $D^{\mu}(x)$ is a local
interpolating field with quantum numbers of the deuteron.

In order to derive the effective Lagrangian of the physical deuteron
field we should integrate over nucleon fields in the one--nucleon loop
approximation [2,14,15]. The one--nucleon loop approximation of
low--energy nuclear forces allows (i) to transfer nucleon flavours
from an initial to a final nuclear state by a minimal way and (ii) to
take into account contributions of nucleon--loop anomalies [28--31],
which are fully defined by one--nucleon loop diagrams [29--31]. It is
well--known that quark--loop anomalies play an important role for the
correct description of strong low--energy interactions of low--lying
hadrons [4--13]. We argue the dominant role of nucleon--loop anomalies
for the correct description of low--energy nuclear forces in nuclear
physics.  We demonstrate below the dominant role of nucleon--loop
anomalies by example of the evaluation of the effective Lagrangian of
the free deuteron field.

The effective Lagrangian of the free deuteron field evaluated in the
one--nucleon loop approximation is defined by [14,15]:
\begin{eqnarray}\label{label3.6}
\hspace{-0.5in}&&\int d^4x\,{\cal L}_{\rm eff}(x) =\int
d^4x\,M^2_0\,D^{\dagger}_{\mu}(x)D^{\mu}(x) \nonumber\\
\hspace{-0.5in}&& - \int d^4x\int\frac{d^4x_1
d^4k_1}{(2\pi)^4}\,e^{\textstyle -ik_1\cdot(x -
x_1)}\,D^{\dagger}_{\mu}(x) D_{\nu}(x_1)\,\frac{g^2_{\rm
V}}{4\pi^2}\,\Pi^{\mu\nu}(k_1; Q),
\end{eqnarray}
where the structure function $\Pi^{\mu\nu}(k_1; Q)$ is given by
\begin{eqnarray}\label{label3.7}
\Pi^{\mu\nu}(k_1; Q)=\int\frac{d^4k}{\pi^2i}{\rm
tr}\Bigg\{\frac{1}{M_{\rm N} - \hat{k} - \hat{Q} -
\hat{k}_1}\gamma^{\mu}\frac{1}{M_{\rm N} - \hat{k} -
\hat{Q}}\gamma^{\nu}\Bigg\}.
\end{eqnarray}
The 4--momentum $Q = a\,k_1$ is an arbitrary shift of momenta of
virtual nucleon fields with an arbitrary parameter $a$. According to
Refs.\,[29,31] the $Q$--dependent parts of one--nucleon loop diagrams
are related to the anomalies of these diagrams, thereby, the correct
evaluation of the $Q$--dependence of one--nucleon loop diagrams is a
great deal of importance in the NNJL model stating a dominant role of
nucleon--loop anomalies. For the evaluation of the $Q$--dependence of
the structure function $\Pi^{\mu\nu}(k_1; Q)$ we apply the procedure
invented by Gertsein and Jackiw [29] (see also [15]):
\begin{eqnarray}\label{label3.8}
&&\Pi^{\mu\nu}(k_1; Q) - \Pi^{\mu\nu}(k_1; 0)
=\int\limits^1_0dx\,\frac{d}{dx}\,\Pi^{\mu\nu}(k_1; x Q)=\nonumber\\
&&=\int\limits^1_0dx\int
\frac{d^4k}{\pi^2i}Q^{\lambda}\frac{\partial}{\partial
k^{\lambda}}{\rm tr}\Bigg\{\frac{1}{M_{\rm N} - \hat{k} - x\hat{Q} -
\hat{k}_1}\gamma^{\mu}\frac{1}{M_{\rm N} - \hat{k} -
x\hat{Q}}\gamma^{\nu}\Bigg\}=\nonumber\\ 
&&= 2 \int\limits^1_0
dx\,\lim_{k\to \infty}\Bigg<\frac{Q\cdot k}{k^2}{\rm tr}\{(M_{\rm N} +
\hat{k} + x\hat{Q} + \hat{k}_1)\gamma^{\mu}(M_{\rm N} + \hat{k} +
x\hat{Q})\gamma^{\nu}\}\Bigg>=\nonumber\\ 
&&=2\,(2Q^{\mu}Q^{\nu} -
Q^2\,g^{\mu\nu}) + 2(k^{\mu}_1Q^{\nu} + k^{\nu}_1Q^{\mu} - k_1\cdot
Q\,g^{\mu\nu})=\nonumber\\ &&= -2\,a(a+1)\,(k^2_1\,g^{\mu\nu} -
2\,k^{\mu}_1k^{\nu}_1).
\end{eqnarray}
Thus, we obtain
\begin{eqnarray}\label{label3.9}
\Pi^{\mu\nu}(k_1; Q) -
\Pi^{\mu\nu}(k_1; 0)= -2\,a(a+1)\,(k^2_1\,g^{\mu\nu} -
2\,k^{\mu}_1k^{\nu}_1).
\end{eqnarray}
We would like to emphasize that the r.h.s. of Eq.(\ref{label3.9}) is
an explicit expression completely defined by high--energy
(short--distance) $N\bar{N}$ fluctuations, since the virtual momentum
$k$ is taken at the limit $k \to \infty$, and related to the anomaly
of the one--nucleon loop diagram with two vector vertices (the
VV--diagram) [29,31].

The structure function $\Pi^{\mu\nu}(k_1; 0)$ has been evaluated in
Refs.\,[14,15] and reads
\begin{eqnarray}\label{label3.10}
\Pi^{\mu\nu}(k_1; 0)=\frac{4}{3}(k^2_1g^{\mu\nu} -
k^{\mu}_1k^{\nu}_1)J_2(M_{\rm
N}) + 2g^{\mu\nu}[J_1(M_{\rm N}) + M^2_{\rm N}J_2(M_{\rm
N})],
\end{eqnarray}
where  $J_1(M_{\rm N})$ and $J_2(M_{\rm N})$ are
the quadratically and logarithmically divergent integrals [14,15]:
\begin{eqnarray}\label{label3.11}
J_1(M_{\rm N})&=&\int\frac{d^4k}{\pi^2i}\frac{1}{M^2_{\rm N} - k^2} =
4 \int\limits^{\textstyle \Lambda_{\rm
D}}_0\frac{d|\vec{k}\,|\vec{k}^{\,2}}{(M^2_{\rm N}
+\vec{k}^{\,2})^{1/2}},\nonumber\\ 
J_2(M_{\rm
N})&=&\int\frac{d^4k}{\pi^2i}\frac{1}{(M^2_{\rm N} - k^2)^2} = 2
\int\limits^{\textstyle \Lambda_{\rm
D}}_0\frac{d|\vec{k}\,|\vec{k}^{\,2}}{(M^2_{\rm N}
+\vec{k}^{\,2})^{3/2}}.
\end{eqnarray}
The cut--off $\Lambda_{\rm D}$ restricts from above 3--momenta of
low--energy fluctuations of virtual neutron and proton fields forming
the physical deuteron [14,15]. As has been shown in Refs.\,[14,15] the cut--off
$\Lambda_{\rm D}$ is much less than the mass of the nucleon, $M_{\rm
N} \gg \Lambda_{\rm D}$ [14,15]. This leads to the relation between the
divergent integrals:
\begin{eqnarray}\label{label3.12}
J_1(M_{\rm N})= 2\,M^2_{\rm N}\,J_2(M_{\rm N}) =
\frac{4}{3}\,\frac{\Lambda^3_{\rm D}}{M_{\rm N}} \sim O(1/N_C)
\end{eqnarray}
which we use below.  Note that in Eq.(\ref{label3.10}) we have taken
into account only the leading terms in the external momentum
expansion, i.e. the $k_1$--expansion [14,15].

The justification of the dominance of the leading order contributions
in expansion in powers of external momenta can be provided in the
large $N_C$ approach to the description of QCD in the
non--perturbative regim. Indeed, in QCD with the $SU(N_C)$ gauge group
at $N_C \to \infty$ the baryon mass is proportional to the number of
quark colours [17]: $M_{\rm N} \sim N_C$.  Since for the derivation of
effective Lagrangians describing the deuteron itself and amplitudes of
processes of low--energy interactions of the deuteron coupled to other
particles all external momenta of interacting particles should be kept
off--mass shell, the masses of virtual nucleon fields taken at $N_C
\to \infty$ are larger compared with external momenta. By expanding
one--nucleon loop diagrams in powers of $1/M_{\rm N}$ we get an
expansion in powers of $1/N_C$. Keeping the leading order in the large
$N_C$ expansion we are leaving with the leading order contributions in
an external momentum expansion. We should emphasize that anomalous
contributions of one--nucleon loop diagrams are defined by the least
powers of an external momentum expansion. This implies that in the
NNJL model effective Lagrangians of low--energy interactions are
completely determined by contributions of one--nucleon loop
anomalies. The divergent contributions having the same order in
momentum expansion are negligible small compared with the anomalous
ones due to the inequality $M_{\rm N} \gg \Lambda_{\rm D}$ and the
limit $N_C \to \infty$. This justifies the application of the
approximation by the leading powers in an external momentum expansion
to the evaluation of the effective Lagrangians of the deuteron coupled
to itself and other fields.

Collecting all pieces we get the structure function $\Pi^{\mu\nu}(k_1;
Q)$ in the form
\begin{eqnarray}\label{label3.13}
\Pi^{\mu\nu}(k_1; Q)&=&\frac{4}{3}(k^2_1g^{\mu\nu} -
k^{\mu}_1k^{\nu}_1)J_2(M_{\rm
N}) + 2g^{\mu\nu}[J_1(M_{\rm N}) + M^2_{\rm N}J_2(M_{\rm
N})]\nonumber\\
&& -  2\,a(a+1)\,(k^2_1\,g^{\mu\nu} -
2\,k^{\mu}_1k^{\nu}_1).
\end{eqnarray}
The effective  Lagrangian of the free deuteron field is then defined by
\begin{eqnarray}\label{label3.14}
&&{\cal L}_{\rm eff}(x)= -\frac{1}{2}\Bigg(-\frac{g^2_{\rm
V}}{2\pi^2}\,a(a+1)+ \frac{g^2_{\rm
V}}{3\pi^2}\,J_2(M_{\rm
N})\Bigg)\,D^{\dagger}_{\mu\nu}(x)D^{\mu\nu}(x)\nonumber\\
&& + \Bigg(M^2_0 - \frac{g^2_{\rm
V}}{2\pi^2}\,[J_1(M_{\rm N}) + M^2_{\rm N}J_2(M_{\rm
N})]\Bigg)\,D^{\dagger}_{\mu}(x)D^{\mu}(x),
\end{eqnarray}
where $D^{\mu\nu}(x)=\partial^{\mu}D^{\nu}(x) -
\partial^{\nu}D^{\mu}(x)$. We have dropped some  contributions
proportional to the total divergence of the deuteron field, since
$\partial_{\mu}D^{\mu}(x) = 0$. For the derivation of
Eq.(\ref{label3.14}) we have used the relation
\begin{eqnarray}\label{label3.15}
&&\int d^4x\int\frac{d^4x_1 d^4k_1}{(2\pi)^4}\,e^{\textstyle
-ik_1\cdot(x - x_1)}\,D^{\dagger}_{\mu}(x)D_{\nu}(x_1)(k^2_1g^{\mu\nu}
- k^{\mu}_1k^{\nu}_1)=\nonumber\\ 
&&= \frac{1}{2}\int
d^4x\,D^{\dagger}_{\mu\nu}(x)D^{\mu\nu}(x).
\end{eqnarray}
In order to get a correct kinetic term of the free deuteron field in
the effective Lagrangian Eq.(\ref{label3.14}) we should set 
\begin{eqnarray}\label{label3.16}
-\frac{g^2_{\rm
V}}{2\pi^2}\,a(a+1)=1.
\end{eqnarray}
Since $a$ is an arbitrary real parameter, the relation
Eq.(\ref{label3.16}) is valid in the case of the existence of real
roots.  For the existence of real roots of Eq.(\ref{label3.16}) the
coupling constant $g_{\rm V}$ should obey the constraint $g^2_{\rm V}
\le 8\pi^2$ that is satisfied by the numerical value $g_{\rm V} =
11.319$ calculated at $N_C = 3$ [15]. Since $g_{\rm V} \sim
O(\sqrt{N_C})$ at $N_C \to \infty$, Eq.(\ref{label3.16}) has real
solutions for any $N_C \ge 3$.

Due to Eq.(\ref{label3.16}) the effective Lagrangian of the free
deuteron field takes the form
\begin{eqnarray}\label{label3.17}
&&{\cal L}_{\rm eff}(x)= -\frac{1}{2}\Bigg(1 + \frac{g^2_{\rm
V}}{3\pi^2}\,J_2(M_{\rm
N})\Bigg)\,D^{\dagger}_{\mu\nu}(x)D^{\mu\nu}(x)\nonumber\\
&& + \Bigg(M^2_0 - \frac{g^2_{\rm
V}}{2\pi^2}\,[J_1(M_{\rm N}) + M^2_{\rm N}J_2(M_{\rm
N})]\Bigg)\,D^{\dagger}_{\mu}(x)D^{\mu}(x).
\end{eqnarray}
By performing the renormalization of the wave function of the deuteron
field [14,15]
\begin{eqnarray}\label{label3.18}
\Bigg(1 + \frac{g^2_{\rm
V}}{3\pi^2}\,J_2(M_{\rm
N})\Bigg)^{1/2}\,D^{\mu}(x) \to D^{\mu}(x)
\end{eqnarray}
and taking into account that $M_{\rm N} \gg \Lambda_{\rm D}$ we arrive
at the effective Lagrangian of the free physical deuteron field
\begin{eqnarray}\label{label3.19}
{\cal L}_{\rm eff}(x)=
-\frac{1}{2}\,D^{\dagger}_{\mu\nu}(x)D^{\mu\nu}(x) + M^2_{\rm
D}\,D^{\dagger}_{\mu}(x)D^{\mu}(x),
\end{eqnarray}
where $M_{\rm D} = M_0 - \varepsilon_{\rm D}$ is the mass of the
physical deuteron field. The binding energy of the deuteron
$\varepsilon_{\rm D}$ reads 
\begin{eqnarray}\label{label3.20}
\varepsilon_{\rm D}= \frac{17}{48}\,\frac{g^2_{\rm
V}}{\pi^2}\,\frac{J_1(M_{\rm N})}{M_{\rm N}}= \frac{17}{18}\,Q_{\rm
D}\,\Lambda^3_{\rm D} \sim O(1/N_C).
\end{eqnarray}
We have used here the relation between divergent integrals
Eq.(\ref{label3.11}) and expressed the phenomenological coupling
constant $g_{\rm V}$ in terms of the electric quadrupole moment of the
deuteron $g^2_{\rm V} = 2\pi^2Q_{\rm D}M^2_{\rm N}$. The dependence of
the physical observable parameter, the binding energy of the deuteron,
on the cut--off $\Lambda_{\rm D}$ is usual for any effective theory
like the NJL model [1--11].

At $N_C \to \infty$ the binding energy of the deuteron behaves like
$O(1/N_C)$ as well as the electric quadrupole moment $Q_{\rm D}$ and
the coupling constant of the phenomenological local four--nucleon
interaction Eq.(\ref{label3.1}). This testifies a self--consistency of
our approach. Really, all parameters of the physical deuteron field
are of the same order according to the large $N_C$ expansion. This
means that the vanishing of the coupling constant of the
phenomenological four--nucleon interaction Eq.(\ref{label3.1}) in the
limit $N_C \to \infty$ entails the vanishing of all low--energy
parameters of the physical deuteron.

\section{Electromagnetic properties of the deuteron}
\setcounter{equation}{0}

\hspace{0.2in} The description of the deuteron as a Cooper np--pair
changes the analysis of the electromagnetic parameters of the deuteron
given in Ref.\,[15], since we do not have more a ``bare'' deuteron
field having the magnetic dipole and electric quadrupole
moment. Therefore, for the Cooper np--pair both the magnetic dipole
and electric quadrupole moments have to be induced fully by the
one--nucleon loop contributions. For the self--consistent description
of the electromagnetic properties of the deuteron we cannot deal with
only the nucleon current $j_{\mu}(x)$ given by Eq.(\ref{label3.2}) and
have to introduce the tensor nucleon current [14,15]
\begin{eqnarray}\label{label4.1}
J^{\mu\nu}(x)= \bar{p^c}(x)\sigma^{\mu\nu}n(x) -
\bar{n^c}(x)\sigma^{\mu\nu}p(x),
\end{eqnarray}
where $\sigma^{\mu\nu}=(\gamma^{\mu}\gamma^{\nu}-
\gamma^{\nu}\gamma^{\mu})/2$.

The local four--nucleon interaction producing the deuteron as a Cooper
np--pair reads now 
\begin{eqnarray}\label{label4.2}
{\cal L}_{\rm int}(x) = - \frac{1}{4M^2_{\rm
N}}\,J^{\dagger}_{\mu}(x)J^{\mu}(x).
\end{eqnarray}
The baryon current $J^{\mu}(x)$ is defined by
\begin{eqnarray}\label{label4.3}
J^{\mu}(x) &=& -i\,g_{\rm V}\,[\bar{p^c}(x)\gamma^{\mu}n(x) -
\bar{n^c}(x)\gamma^{\mu}p(x)]\nonumber\\
&&-\,\frac{g_{\rm T}}{M_{\rm
N}}\,\partial_{\nu}[\bar{p^c}(x)\sigma^{\nu\mu}n(x) -
\bar{n^c}(x)\sigma^{\nu\mu}p(x)],
\end{eqnarray}
where $g_{\rm T}$ is a dimensionless phenomenological coupling
constant [15]. The contribution of the tensor nucleon current looks
like the next--to--leading term in the long--wavelength
expansion\footnote{Due to proportionality $M_{\rm N} \sim N_C$ this
expansion is related to the large $N_C$ expansion.} of an effective
low--energy four--nucleon interaction.

The effective Lagrangian of the np--system unstable under creation of
the Cooper np--pair with quantum numbers of the deuteron
is then defined
\begin{eqnarray}\label{label4.4}
{\cal L}^{\rm np}(x)&=&\bar{n}(x)\,(i\gamma^{\mu}\partial_{\mu} - M_{\rm
N})n(x)+ \bar{p}(x)(i\gamma^{\mu}\partial_{\mu} -
M_{\rm N})p(x)  \nonumber\\ 
&&- \frac{1}{4M^2_{\rm
N}}J^{\dagger}_{\mu}(x)J^{\mu}(x).
\end{eqnarray}
The linearalized version of the effective Lagrangian
Eq.(\ref{label4.4}) containing the interpolating local deuteron field
reads
\begin{eqnarray}\label{label4.5}
\hspace{-0.5in}{\cal L}^{\rm np}(x)&\to& \bar{n}(x)(i\gamma^{\mu}\partial_{\mu} - M_{\rm
N})n(x) + \bar{p}(x)(i\gamma^{\mu}\partial_{\mu} - M_{\rm
N})p(x)\nonumber\\
\hspace{-0.5in}&&  + M^2_0D^{\dagger}_{\mu}(x)D^{\mu}(x) + g_{\rm V}j^{\dagger}_{\mu}(x)D^{\mu}(x) + g_{\rm
V}j^{\mu}(x)D^{\dagger}_{\mu}(x)\nonumber\\
\hspace{-0.5in}&& + \frac{g_{\rm T}}{M_0}J^{\dagger}_{\mu\nu}(x)
D^{\mu\nu}(x) + \frac{g_{\rm T}}{M_0}J^{\mu\nu}(x)
D^{\dagger}_{\mu\nu}(x),
\end{eqnarray}
where $M_0 = 2M_{\rm N}$, $D^{\mu}(x)$ is a local interpolating
field with quantum numbers of the deuteron and $D^{\mu\nu}(x) =
\partial^{\mu} D^{\nu}(x) - \partial^{\nu} D^{\mu}(x)$.

The interactions with the tensor current give the contributions only
to the divergent part of the effective Lagrangian of the free deuteron
field determined now by
\begin{eqnarray}\label{label4.6}
\hspace{-0.3in}&&{\cal L}_{\rm eff}(x)=
-\frac{1}{2}\Bigg(-\frac{g^2_{\rm V}}{2\pi^2}a(a+1)+ \frac{g^2_{\rm
V} + 6g_{\rm V}g_{\rm T} + 3g^2_{\rm T}}{3\pi^2}J_2(M_{\rm
N})\Bigg)D^{\dagger}_{\mu\nu}(x)D^{\mu\nu}(x)\nonumber\\
\hspace{-0.3in}&&\hspace{0.7in} + \Bigg(M^2_0 - \frac{g^2_{\rm
V}}{2\pi^2}[J_1(M_{\rm N}) + M^2_{\rm N}J_2(M_{\rm
N})]\Bigg)D^{\dagger}_{\mu}(x)D^{\mu}(x).
\end{eqnarray}
The one--nucleon loop diagrams defining the effective Lagrangian
Eq.(\ref{label4.6}) are depicted in Fig.1.

Due to the relation Eq.(\ref{label3.16}) the effective Lagrangian of
the free deuteron field Eq.(\ref{label4.6}) takes the form
\begin{eqnarray}\label{label4.7}
\hspace{-0.3in}&&{\cal L}_{\rm eff}(x)=
-\frac{1}{2}\Bigg(1 + \frac{g^2_{\rm
V} +  6g_{\rm V}g_{\rm T} + 3g^2_{\rm T}}{3\pi^2}J_2(M_{\rm
N})\Bigg)D^{\dagger}_{\mu\nu}(x)D^{\mu\nu}(x)\nonumber\\
\hspace{-0.3in}&&\hspace{0.7in} + \Bigg(M^2_0 - \frac{g^2_{\rm
V}}{2\pi^2}[J_1(M_{\rm N}) + M^2_{\rm N}J_2(M_{\rm
N})]\Bigg)D^{\dagger}_{\mu}(x)D^{\mu}(x).
\end{eqnarray}
After the renormalization of the wave function of the deuteron field
we arrive at the effective Lagrangian defined by Eq.(\ref{label3.19})
with the binding energy of the deuteron depending on $g_{\rm V}$ and
$g_{\rm T}$
\begin{eqnarray}\label{label4.8}
\varepsilon_{\rm D} &=& \frac{17}{48}\frac{g^2_{\rm
V}}{\pi^2}\,\frac{J_1(M_{\rm N})}{M_{\rm N}}\Bigg( 1 +
\frac{48}{17}\,\frac{g_{\rm T}}{g_{\rm V}} +
\frac{24}{17}\,\frac{g^2_{\rm T}}{g^2_{\rm V}}\Bigg) =\nonumber\\
&=&\frac{17}{18}\,Q_{\rm D}\,\Lambda^3_{\rm D}\,\Bigg( 1 +
\frac{48}{17}\,\frac{g_{\rm T}}{g_{\rm V}} +
\frac{24}{17}\,\frac{g^2_{\rm T}}{g^2_{\rm V}}\Bigg),
\end{eqnarray}
where we have used the relation between divergent integrals
Eq.(\ref{label3.11}) and expressed the phenomenological coupling
constant $g_{\rm V}$ in terms of the electric quadrupole moment of the
deuteron $g^2_{\rm V} = 2\pi^2Q_{\rm D}M^2_{\rm N}$. In order to make
the prediction for the binding energy much more definite we have to
know the relation between the phenomenological coupling constants
$g_{\rm V}$ and $g_{\rm T}$. For this aim we suggest to consider the
electromagnetic properties of the deuteron.

Including the electromagnetic field by a minimal way $\partial_{\mu}
\to \partial_{\mu} + i\,e\,A_{\mu}(x)$, where $e$ and $A_{\mu}(x)$ are
the electric charge of the proton and the electromagnetic potential we
bring up the linearalized version of the Lagrangian
Eq.(\ref{label4.5}) to the form 
\begin{eqnarray}\label{label4.9}
\hspace{-0.5in}&&{\cal L}^{\rm np}(x) \to {\cal L}^{\rm np}_{\rm
ELM}(x) =\nonumber\\
\hspace{-0.5in}&&= \bar{n}(x)(i\gamma^{\mu}\partial_{\mu} - M_{\rm
N})n(x) + \bar{p}(x)(i\gamma^{\mu}\partial_{\mu} - M_{\rm
N})p(x) + M^2_0D^{\dagger}_{\mu}(x)D^{\mu}(x)\nonumber\\
\hspace{-0.5in}&& + g_{\rm V}j^{\dagger}_{\mu}(x)D^{\mu}(x) + g_{\rm
V}j^{\mu}(x)D^{\dagger}_{\mu}(x) + \frac{g_{\rm
T}}{M_0}J^{\dagger}_{\mu\nu}D^{\mu\nu}(x) + \frac{g_{\rm
T}}{M_0}J^{\mu\nu}D^{\dagger}_{\mu\nu}(x)\nonumber\\ 
\hspace{-0.5in}&&-\,e\,\bar{p}(x)\gamma^{\mu}p(x)\,A_{\mu}(x) 
- i\,e\,\frac{g_{\rm T}}{M_0}J^{\dagger}_{\mu\nu}(x)(A^{\mu}(x)D^{\nu}(x) -
A^{\nu}(x)D^{\mu}(x))\nonumber\\
\hspace{-0.5in}&& + i\,e\,\frac{g_{\rm
T}}{M_0}J^{\mu\nu}(x)(A_{\mu}(x)D_{\nu}(x) - A^{\nu}(x)D^{\mu}(x)).
\end{eqnarray}
By using this Lagrangian we should calculate fully all contributions
to the effective Lagrangian of the deuteron coupled to an external
electromagnetic field. These are the effective Lagrangians of the
Corben--Schwinger [18] and the Aronson [19] type defining at field
theoretic level the magnetic dipole and the electric quadrupole moment
of the deuteron, and the effective interactions which can be
identified with the contributions caused by the minimal inclusion of
the electromagnetic field $\partial_{\mu} D_{\nu}(x) \to
(\partial_{\mu} + i\,e\,A_{\mu}(x)) D_{\nu}(x)$.

\subsection{The phenomenological Corben--Schwinger interaction}

\hspace{0.2in} The one--nucleon loop diagrams defining in the NNJL
model effective electromagnetic interactions of the deuteron
linear in electric charge $e$ induced by the Lagrangian
Eq.(\ref{label4.9}) are depicted in Fig.2. One can show that in the
$1/M_{\rm N}$ expansion corresponding to the large $N_C$ expansion due
to the proportionality $ M_{\rm N} \sim N_C$ [17] the one--nucleon
loop diagrams in Fig.2a and 2b are divergent. Therefore, due to
Eq.(\ref{label3.12}) at leading order in the large $N_C$ expansion the
contributions of these diagrams can be neglected with respect to the
contributions of the diagrams in Fig.2c and 2d defining the
phenomenological Lagrangians of the Corben--Schwinger, ${\cal L}_{\rm
CS}(x)$, and the Aronson, ${\cal L}_{\rm A}(x)$, type, respectively,
in terms of the nucleon--loop anomalies [15].

The effective Lagrangian of the diagram in Fig.2c is defined by [15]
\begin{eqnarray}\label{label4.10}
\hspace{-0.5in}\int d^4 x\,{\cal L}_{\rm Fig.2c}(x)&=&\int d^4x\,
\int \frac{d^4 x_1 d^4 k_1}{(2\pi)^4}\, \frac{d^4x_2 d^4 k_2}{(2
\pi)^4}\,D_{\beta}(x)\,D^{\dagger}_{\alpha}(x_1)\,A_{\mu}(x_2)
\nonumber\\ 
\hspace{-0.5in}&&\times \, e^{- i\,k_1\cdot x_1}\,e^{-
i\,k_2\cdot x_2}\,e^{i\,(k_1 + k_2)\cdot x}\,\frac{e g^2_{\rm V}}{4
\pi^2}\,{\cal J}^{\beta\alpha\mu}(k_1, k_2; Q).
\end{eqnarray}
In the one--nucleon loop approximation the structure function ${\cal
J}^{\beta\alpha\mu}(k_1, k_2; Q)$ is given by the momentum integral
[15]
\begin{eqnarray}\label{label4.11}
\hspace{-0.7in}&&{\cal J}^{\beta\alpha\mu}(k_1, k_2; Q) = 
\int\frac{d^4\,k}{\pi^2\,i}\nonumber\\
\hspace{-0.7in}&&\times\,{\rm
tr}\Bigg\{\gamma^{\beta} \frac{1}{M_{\rm
N}-\hat{k}-\hat{Q}} \gamma^{\alpha} \frac{1}{M_{\rm
N}-\hat{k}-\hat{Q}-\hat{k}_1} \gamma^{\mu} \frac{1}{M_{\rm
N}-\hat{k}-\hat{Q}-\hat{k}_1-\hat{k}_2}\Bigg\}.
\end{eqnarray}
The 4--vector $Q\,=\,a\,k_1 + b\,k_2$, where $a$ and $b$ are arbitrary
parameters, displays the dependence of the $k$ integral in
(\ref{label4.11}) on a shift of a virtual momentum. According to
Refs.[29,31] a $Q$--dependent part of an one--nucleon loop diagram is
related to the anomaly of this diagram. Therefore, the evaluation of
the $Q$--dependence of the one--nucleon loop diagram should play an
important role in the NNJL model. For the evaluation of the
$Q$--dependence of the structure function Eq.(\ref{label4.11}) we
apply the method invented by Gertsein and Jackiw [29] and consider the
following difference of momentum integrals
\begin{eqnarray}\label{label4.12}
\delta\,{\cal J}^{\beta\alpha\mu}(k_1, k_2; Q)\,=\,{\cal
J}^{\beta\alpha\mu}(k_1, k_2; Q\,) - {\cal
J}^{\beta\alpha\mu}(k_1, k_2; 0)
\end{eqnarray}
In accordance with the Gertsein--Jackiw method the difference
(\ref{label4.12}) can be represented by the integral
\begin{eqnarray}\label{label4.13}
&&\delta\,{\cal J}^{\beta\alpha\mu}(k_1, k_2; Q)\,=\,\int^{1}_{0} d x\,\frac{d}{d\,x}{\cal J}^{\beta\alpha\mu}(k_1, k_2; x\,Q)\,=\nonumber\\
&&= \int^{1}_{0}\,d\,x\,\int\frac{d^4\,k}{\pi^2\,i}\,Q^{\lambda}\,\frac{\partial}{\partial\,k^{\lambda}}\,{\rm tr}
\Bigg\{\gamma^{\beta}\,\frac{1}{M_{\rm N}-\hat{k}-x\hat{Q}}\,
\gamma^{\alpha}\,\frac{1}{M_{\rm N}-\hat{k}-x\hat{Q}-\hat{k}_1}\,
\gamma^{\mu}\,\nonumber\\
&&\times \,\frac{1}{M_{\rm N}-\hat{k}-x\hat{Q}-\hat{k}_1-\hat{k}_2}\Bigg\}.
\end{eqnarray}
This shows that the contribution of the $Q$--dependent part of the
structure function Eq.(\ref{label4.11}) is just the surface
term. Following Gertsein and Jackiw [29] and evaluating the integral
over $k$ symmetrically we obtain
\begin{eqnarray}\label{label4.14}
&&\delta\,{\cal J}^{\beta\alpha\mu}(k_1, k_2;
Q)\,=\,-\,2\,\int^{1}_{0} d x\,\lim_{k\to \infty}\Bigg<\frac{Q\cdot
k}{k^4}{\rm tr}\{\gamma^{\beta}\,(M_{\rm
N}+\hat{k}+x\hat{Q})\,\gamma^{\alpha}\,\times \nonumber\\ &&\times
(M_{\rm N}+\hat{k}+x\hat{Q}+\hat{k}_1)\,\gamma^{\mu}\,(M_{\rm
N}+\hat{k}+x\hat{Q}+\hat{k}_1+\hat{k}_2)\}\Bigg>.
\end{eqnarray}
The brackets $<\ldots>$ mean the averaging over $k$ directions. Due to
the limit $k\to \infty$ we can neglect all momenta with respect to
$k$.
\begin{eqnarray}\label{label4.15}
&&\delta\,{\cal J}^{\beta\alpha\mu}(k_1, k_2; Q)\,=\,-\,2\,\lim_{R\to
\infty}\Bigg<\frac{Q\cdot k}{k^4}{\rm tr}\{\gamma^{\beta} \hat{k}
\gamma^{\alpha} \hat{k} \gamma^{\mu} \hat{k}\}\Bigg>.
\end{eqnarray}
Averaging over $k$--directions
\begin{eqnarray}\label{label4.16}
\lim_{k\to
\infty}\frac{k^{\lambda}k^{\varphi}k^{\omega}k^{\rho}}{k^4}\,=\,
\frac{1}{24}\,(g^{\lambda\varphi}
g^{\omega\rho} + g^{\lambda\omega} g^{\varphi \rho} + g^{\lambda\rho}
g^{\varphi\omega})
\end{eqnarray}
we obtain
\begin{eqnarray}\label{label4.17}
&&\delta\,{\cal J}^{\beta\alpha\mu}(k_1, k_2;
Q)\,=\,-\,\frac{1}{12}\,{\rm tr}(\gamma_{\lambda} \gamma^{\beta}
\gamma^{\lambda} \gamma^{\alpha} \hat{Q} \gamma^{\mu} + \gamma^{\beta}
\gamma_{\lambda} \gamma^{\alpha} \gamma^{\lambda} \hat{Q} \gamma^{\mu}
\nonumber\\
&&+\,\gamma_{\beta} \hat{Q} \gamma^{\alpha} \gamma_{\lambda}
\gamma^{\mu} \gamma^{\lambda}) = \frac{2}{3}\,(Q^{\alpha} g^{\beta\mu}
+ Q^{\beta} g^{\mu\alpha} + Q^{\mu} g^{\alpha\beta}).
\end{eqnarray}
Our result Eq.(\ref{label4.17}) agrees with the statement by Gertsein
and Jackiw [29] that the $Q$--dependence of one--nucleon loop
diagrams, i.e. the anomaly of the one--nucleon loop diagram, is fully
defined by the surface behavior of the integrand of the momentum
integral at a virtual momentum going to infinity, $k \to \infty$. This
relates the anomalies of the one--nucleon loop diagrams with
contributions of high--energy (short--distance) fluctuations of
virtual nucleon and anti--nucleon fields, i.e. the $N\bar{N}$
fluctuations.

Now we can proceed to the evaluation of ${\cal
J}^{\beta\alpha\mu}(k_1, k_2; Q)$. In order to pick up the
contribution of the $Q$--dependent part one cannot apply the Feynman
method of the evaluation of momentum integrals like
(\ref{label4.11}). This method involves the mergence of the factors in
the denominator with the subsequent shift of a virtual momentum. On
this way one can lose the $Q$--dependence by virtue of the shift at
the intermediate stage. Thereby, we have to evaluate the integral over
$k$ without any intermediate shifts.

One can make this by applying a momentum expansion related to the
$1/M_{\rm N}$ expansion or that is the same the large $N_C$ expansion
[17] and keeping only the leading terms.
\begin{eqnarray}\label{label4.18}
\hspace{-0.3in}&&{\cal J}^{\beta\alpha\mu}(k_1, k_2; Q)\,=\nonumber\\
\hspace{-0.3in}&&=\,\int\frac{d^4\,k}{\pi^2 i}\,{\rm
tr}\Bigg\{\gamma^{\beta}\,\frac{M_{\rm N}+\hat{k}+\hat{Q}}{M^2_{\rm
N}-k^2}\,\Bigg[1 + \frac{2\,k\cdot Q}{M^2_{\rm N}-k^2}\Bigg]\,
\gamma^{\alpha}\,\frac{M_{\rm N}+\hat{k}+\hat{Q}+\hat{k}_1}{M^2_{\rm
N}-k^2}\,\times \nonumber\\ &&\times\,\Bigg[1 + \frac{2\,k\cdot
(Q+k_1)}{M^2_{\rm N}-k^2}\Bigg]\,\gamma^{\mu}\,\frac{M_{\rm
N}+\hat{k}+\hat{Q}+\hat{k}_1+\hat{k}_2}{M^2_{\rm N}-k^2}\nonumber\\
\hspace{-0.3in}&&\times\,\Bigg[1 + \frac{2\,k\cdot (Q+k_1+k_2)}{M^2_{\rm
N}-k^2}\Bigg]\Bigg\} =\nonumber\\
\hspace{-0.3in}&&=\,\int\frac{d^4\,k}{\pi^2\,i}\,\frac{1}{(M^2_{\rm N}-k^2)^{3}}{\rm
tr}\,\{M^2_{\rm
N}\gamma^{\beta}(\hat{k}+\hat{Q})\gamma^{\alpha}\gamma^{\mu}+M^2_{\rm
N}\gamma^{\beta}\gamma^{\alpha}(\hat{k}+\hat{Q}+\hat{k}_1)\gamma^{\mu}
\nonumber\\ 
\hspace{-0.3in}&&+M^2_{\rm N}\gamma^{\beta}\gamma^{\alpha}
\gamma^{\mu}(\hat{k}+\hat{Q}+\hat{k}_1+\hat{k}_2)+
\gamma^{\beta}(\hat{k}+\hat{Q})\gamma^{\alpha}
(\hat{k}+\hat{Q}+\hat{k}_1)\nonumber\\ 
\hspace{-0.3in}&&\times \gamma^{\mu}(\hat{k}+\hat{Q}+
\hat{k}_1+\hat{k}_2)\}\Bigg[1 + \frac{2\,k\cdot
(3Q+2k_1+k_2)}{M^2_{\rm N}-k^2}\Bigg]\,=\nonumber\\
\hspace{-0.3in}&&=\,\frac{1}{2}\int\frac{d^4\,k}{\pi^2\,i}
\Bigg[\frac{1}{(M^2_{\rm
N}-k^2)^2}+\frac{M^2_{\rm N}}{(M^2_{\rm N}-k^2)^{3}}\Bigg]{\rm
tr}\,\{\gamma^{\beta}\hat{Q}\gamma^{\alpha}\gamma^{\mu}+
\gamma^{\beta}\gamma^{\alpha}(\hat{Q}+\hat{k}_1)\gamma^{\mu}\nonumber\\
\hspace{-0.3in}&&+\gamma^{\beta}\gamma^{\alpha}\gamma^{\mu}(\hat{Q}+\hat{k}_1+\hat{k}_2)\}
\nonumber\\ 
\hspace{-0.3in}&&+\,2\,\int\frac{d^4\,k}{\pi^2\,i}\frac{k\cdot
(3Q+2k_1+k_2)}{(M^2_{\rm N}-k^2)^{3}}{\rm tr}\,\{M^2_{\rm
N}(\gamma^{\beta}\hat{k}\gamma^{\alpha}\gamma^{\mu}+\gamma^{\beta}
\gamma^{\alpha}\hat{k}\gamma^{\mu}+\gamma^{\beta}\gamma^{\alpha}
\gamma^{\mu}\hat{k})\nonumber\\
\hspace{-0.3in}&&+\gamma^{\beta}\hat{k}\gamma^{\alpha}
\hat{k}\gamma^{\mu}\hat{k}\}
\,=\,{\cal J}^{\beta\alpha\mu}_{(1)}(k_1, k_2; Q) + {\cal
J}^{\beta\alpha\mu}_{(2)}(k_1, k_2; Q).
\end{eqnarray}
For the evaluation of ${\cal
J}^{\beta\alpha\mu}_{(1)}(k_1, k_2; Q)$ it is sufficient to
calculate the trace of the Dirac matrices and integrate over $k$
\begin{eqnarray}\label{label4.19}
&&{\cal J}^{\beta\alpha\mu}_{(1)}(k_1, k_2; Q)\,=\,[1 + 2\,J_2(M_{\rm
N})]\, [(Q + 2\,k_1 + k_2)^{\alpha}g^{\beta\,\mu}\nonumber\\
&&+\,(Q + 2\,k_1 + k_2)^{\beta}g^{\mu\,\alpha} + (Q + 2\,k_1 +
k_2)^{\mu}g^{\alpha\,\beta}\nonumber\\ 
&&-\,2\,(k_1 +
k_2)^{\alpha}\,g^{\beta\,\mu} - 2\,k^{\beta}_1\,g^{\mu\,\alpha}],
\end{eqnarray}
where $J_2(M_{\rm N})$ describes a divergent contribution depending on
the cut--off $\Lambda_{\rm D}$. Due to inequality $M_{\rm N}\gg
\Lambda_{\rm D}$ we can neglect $J_2(M_{\rm N})$ with respect to the
convergent contribution. This corresponds too the accounting for the
leading contributions in the large $N_C$ expansion. Indeed, according
to Eq.(\ref{label3.12}) the contribution of divergent integrals is of
order $O(1/N_C)$ relative to the convergent ones.

For the evaluation of ${\cal J}^{\beta\alpha\mu}_{(2)}(k_1, k_2; Q)$
it is convenient, first, to integrate over $k$ directions and then to
calculate the trace over Dirac matrices.  This gives
\begin{eqnarray}\label{label4.20}
&&{\cal J}^{\beta\alpha\mu}_{(2)}(k_1, k_2; Q)\,=\nonumber\\
&&=\,\int\frac{d^4\,k}{\pi^2\,i}\,\Bigg[\frac{1}{2}\frac{M^2_{\rm
N}\,k^2}{(M^2_{\rm N}-k^2)^4}-\frac{1}{6}\frac{k^4}{(M^2_{\rm
N}-k^2)^4}\Bigg]\,(3\,Q + k_1 + k_2)_{\lambda}\nonumber\\
&&\quad\quad\times {\rm
tr}\,(\gamma^{\beta} \gamma^{\lambda} \gamma^{\alpha} 
\gamma^{\mu} + \gamma^{\beta} \gamma^{\alpha} \gamma^{\lambda} 
\gamma^{\mu} + \gamma^{\beta} \gamma^{\alpha} \gamma^{\mu} 
\gamma^{\lambda})\,=\nonumber\\
&&=\,-\,\frac{1}{9}[1 + 6\,J_2(M_{\rm N})]\,[(3\,Q + 2\,k_1 +
k_2)^{\alpha}g^{\beta\,\mu} + (3\,Q + 2\,k_1 +
k_2)^{\beta}g^{\mu\,\alpha}\nonumber\\ 
&&\quad + (3\,Q + 2\,k_1 +
k_2)^{\mu}g^{\alpha\,\beta}].
\end{eqnarray}
Here we have used the integrals
\begin{eqnarray}\label{label4.21}
\int\,\frac{d^4 k}{\pi^2\,i}\,\frac{1}{(M^2_{\rm
N}-k^2)^{3}}&=&\frac{1}{2\,M^2_{\rm N}},\nonumber\\
\int\,\frac{d^4 k}{\pi^2\,i}\,\frac{1}{(M^2_{\rm
N}-k^2)^4}&=&\frac{1}{6\,M^4_{\rm N}}.
\end{eqnarray}
Summing up the contributions  we obtain
\begin{eqnarray}\label{label4.22}
&&{\cal J}^{\beta\alpha\mu}(k_1, k_2;
Q)\,=\,\frac{2}{3}\,(Q^{\alpha}\,g^{\beta\,\mu} +
Q^{\beta}\,g^{\mu\,\alpha} + Q^{\mu}\,g^{\alpha\,\beta}) +
\frac{8}{9}\,[1 + \frac{3}{2}\,J_2(M_{\rm N})]\nonumber\\
&&\times [(2\,k_1 + k_2)^{\alpha}\,g^{\beta\,\mu} + (2\,k_1 +
k_2)^{\beta}\,g^{\mu\,\alpha} + (2\,k_1 +
k_2)^{\mu}\,g^{\alpha\,\beta}]\nonumber\\ 
&&+\,[1 + 2\,J_2(M_{\rm
N})][-\,2\,(k_1 +
k_2)^{\alpha}\,g^{\beta\mu} - 2\,k_1^{\beta}\,g^{\mu\alpha}].
\end{eqnarray}
It is seen that the $Q$--dependence coincides with that obtained by
means of the Gertsein--Jackiw method (\ref{label4.17}). Due to the
arbitrariness of $Q$ we can absorb by the $Q$--term the terms having
the same Lorentz structure. This brings up the r.h.s. of
(\ref{label4.22}) to the form
\begin{eqnarray}\label{label4.23}
{\cal J}^{\beta\alpha\mu}(k_1, k_2; Q)&=& \frac{2}{3}\,(Q^{\alpha}
g^{\beta\mu} + Q^{\beta} g^{\mu\alpha} +
Q^{\mu}g^{\alpha\,\beta})\nonumber\\ 
&&+\,[-\,2\,(k_1 + k_2)^{\alpha}
g^{\beta\mu} - 2\,k_1^{\beta} g^{\mu\alpha}].
\end{eqnarray}
Also we have dropped here the divergent contribution. This
approximation is valid due to the inequality $M_{\rm N}\gg
\Lambda_{\rm D}$ and at leading order in the large $N_C$ expansion.

The effective Lagrangian ${\cal L}_{\rm Fig.2c}(x)$ determined by the
structure function (\ref{label4.23}) reads
\begin{eqnarray}\label{label4.24}
&&{\cal L}_{\rm Fig.2c}(x)\,=\,i\,e\,\frac{g^2_{\rm V}}{6\pi^2}[(3 -
a)\,\partial^{\mu}D^{\dagger}_{\mu}(x)\,D_{\nu}(x)\,
A^{\nu}(x)\nonumber\\ &&-\,(3 -
a)\,D^{\dagger}_{\mu}(x)\,\partial^{\nu}D_{\nu}(x)\,A^{\mu}(x)
-
b\,D^{\dagger}_{\mu}(x)\,D_{\nu}(x)\,\partial^{\mu}\,A^{\nu}(x)
\nonumber\\
&&-\,(b -
a)\,D^{\dagger}_{\mu}(x)\,D_{\nu}(x)\,\partial^{\nu}\,A^{\mu}(x)
\nonumber\\
&&-\,(a -
b)\,\partial^{\nu}\,D^{\dagger}_{\mu}(x)\,D^{\mu}(x)\,A_{\nu}(x)
+
b\,D^{\dagger}_{\mu}(x)\,\partial^{\nu}\,D^{\mu}(x)\,A_{\nu}(x)
\nonumber\\
&&+\,3\,D^{\dagger}_{\mu}(x)\,D_{\nu}(x)\,(\partial^{\mu}\,
A^{\nu}(x) - \partial^{\nu}\,A^{\mu}(x))].
\end{eqnarray}
Due to the constraints $\partial^{\mu}D^{\dagger}_{\mu}(x)\,=\,
\partial^{\mu}D_{\mu}(x)\,=\,0$ some terms in the Lagrangian
(\ref{label4.24}) can be dropped out. This gives
\begin{eqnarray}\label{label4.25}
\hspace{-0.5in}&&{\cal L}_{\rm Fig.2c}(x)\,=\,\nonumber\\
\hspace{-0.5in}&&=i\,e\,\frac{g^2_{\rm V}}{6\pi^2}
[-\,b\,D^{\dagger}_{\mu}(x)\,D_{\nu}(x)\,\partial^{\mu}
A^{\nu}(x)\,-\,(b - a)\,D^{\dagger}_{\mu}(x)\,D_{\nu}(x)\,\partial^{\nu}
A^{\mu}(x)\nonumber\\
\hspace{-0.5in}&&-\,(a - b)\,\partial^{\nu}D^{\dagger}_{\mu}(x)\,D^{\mu}(x)\,
A_{\nu}(x) + b\,D^{\dagger}_{\mu}(x)\,\partial_{\nu} D^{\mu}(x)\,
A^{\nu}(x)\nonumber\\
\hspace{-0.5in}&&+\,3\,D^{\dagger}_{\mu}(x)\,D_{\nu}(x)\,(\partial^{\mu}
A^{\nu}(x)\,-\,\partial^{\nu}A^{\mu}(x))].
\end{eqnarray}
By using the relations $\partial_{\nu} D^{\dagger}_{\mu}(x) =
D^{\dagger}_{\nu\mu}(x) +
\partial_{\mu}D^{\dagger}_{\nu}(x)$ and
$\partial_{\nu} D_{\mu}(x) = D_{\nu\mu}(x) +
\partial_{\mu} D_{\nu}(x)$ we can rewrite the Lagrangian
(\ref{label4.25}) as follows
\begin{eqnarray}\label{label4.26}
\hspace{-0.3in}&&{\cal L}_{\rm Fig.2c}(x) = \nonumber\\
\hspace{-0.3in}&&=i\,e\,\frac{g^2_{\rm V}}{6\pi^2}
[-\,(a - b)\,D^{\dagger}_{\nu\mu}(x)\,A^{\nu}(x)\,
D^{\mu}(x) + b\,D^{\nu\mu}(x)\,A_{\nu}(x)\,
D^{\dagger}_{\mu}(x)\nonumber\\
\hspace{-0.3in}&&-\,(a - b)\,\partial_{\nu} D^{\dagger}_{\mu}(x)\,
D^{\nu}(x)\,A^{\mu}(x) + b\,D^{\dagger}_{\mu}(x)\,
\partial^{\mu} D_{\nu}(x)\,A^{\nu}(x)\nonumber\\
\hspace{-0.3in}&&-\,b\,D^{\dagger}_{\mu}(x)\,D_{\nu}(x)\,
\partial^{\mu} 
A^{\nu}(x) - (b - a)\,D^{\dagger}_{\mu}(x)\,D_{\nu}(x)\,
\partial^{\nu} A^{\mu}(x)\nonumber\\
\hspace{-0.3in}&&+\,3\,D^{\dagger}_{\mu}(x)\,D_{\nu}(x)\,
(\partial^{\mu} 
A^{\nu}(x)\,-\,\partial^{\nu} A^{\mu}(x))].
\end{eqnarray}
The subsequent transformations we perform by applying the identity
\begin{eqnarray}\label{label4.27}
&&\partial^{\nu} D^{\dagger}_{\mu}(x)\,D_{\nu}(x)
\,A^{\mu}(x)\,-\,D^{\dagger}_{\mu}(x)\,\partial^{\mu} 
D_{\nu}(x)\,A^{\nu}(x)=\nonumber\\
&&= D^{\dagger}_{\mu}(x)\,D_{\nu}(x)\,(\partial^{\mu}
A^{\nu}(x)
- \partial^{\nu}\,A^{\mu}(x))
\end{eqnarray}
being valid up to the contribution of a total divergence which can be
omitted. Setting $a = 2\,b$ we represent the effective Lagrangian
(\ref{label4.27}) in the irreducible form
\begin{eqnarray}\label{label4.28}
{\cal L}_{\rm Fig.2c}(x)&=&i\,e\,\frac{g^2_{\rm
V}}{6\pi^2}[b\,D^{\dagger}_{\mu\nu}(x)\,A^{\nu}(x)\,D^{\mu}(x)
-
b\,D^{\mu\nu}(x)\,A_{\nu}(x)\,D^{\dagger}_{\mu}(x)]\nonumber\\
&+& i\,e\,\frac{g^2_{\rm
V}}{6\pi^2}(-\,2\,b +
3)\,D^{\dagger}_{\mu}(x)\,D_{\nu}(x)\,F^{\mu\nu}(x),
\end{eqnarray}
where $F^{\mu\nu}(x) = \partial^{\mu}A^{\nu}(x) -
\partial^{\nu}A^{\mu}(x)$ is the electromagnetic field strength
tensor. Then, first two terms define the finite contributions to the
renormalization constant of the wave function of the deuteron, whereas
the last term coincides with the well--known phenomenological
interaction ${\cal L}_{\rm CS}(x)$ introduced by Corben and Schwinger
[18]
\begin{eqnarray}\label{label4.29}
{\cal L}_{\rm CS}(x) = i\,e\,\frac{g^2_{\rm
V}}{6\pi^2}\,(-\,2\,b +
3)\,D^{\dagger}_{\mu}(x)\,D_{\nu}(x)\,F^{\mu\nu}(x)
\end{eqnarray}
for the description of the charged vector field coupled to an external
electromagnetic field.

Thus, at leading order in the large $N_C$ expansion the anomaly of the
one--nucleon loop triangle VVV--diagram with vector (V) vertices
defines fully the effective Lagrangian of the Corben--Schwinger type
describing the deuteron coupled to an external electromagnetic
field. In turn, the finite the contributions to the renormalization
constant of the wave function of the deuteron we will identify below
with those induced by a minimal inclusion of the electromagnetic
interaction: $\partial_{\mu} D_{\nu}(x) \to (\partial_{\mu}+
i\,e\,A_{\nu}(x)) D_{\nu}(x)$. These terms are important for the
correct definition of the effective Lagrangian of the deuteron coupled
to an electromagnetic field.

\subsection{The phenomenological Aronson interaction}

\hspace{0.2in} The effective Lagrangian described by the diagram in
Fig.2d is defined by [15]
\begin{eqnarray}\label{label4.30}
\hspace{-0.5in}&&\int d^4 x\,{\cal L}_{\rm Fig.2d}(x)\,=\, \int d^4 x
\int \frac{d^4 x_1\,d^4 k_1}{(2\pi)^4}\, \frac{d^4 x_2\,d^4
k_2}{(2\,\pi)^4}\,D_{\alpha\,\beta}(x)\,
D^{\dagger}_{\mu\,\nu}(x_1)\,A_{\lambda}(x_2)\nonumber\\
\hspace{-0.5in}&&\times \, e^{-\,i\,k_1\cdot x_1}\,e^{-\,i\,k_2\cdot x_2}
\,e^{i\,(k_1\,+\,k_2)\cdot x}\,(-\,e)\,
\frac{g^{2}_{\rm T}}{4\pi^{2}}\,\frac{1}{M^{2}_{\rm D}}\,
{\cal J}^{\alpha\beta\mu\nu\lambda}(k_1, k_2; Q).
\end{eqnarray}
In the structure function ${\cal
J}^{\alpha\beta\mu\nu\lambda}(k_1, k_2; Q)$ is represented by the
following momentum integral

\begin{eqnarray}\label{label4.31}
\hspace{-0.3in}&&{\cal J}^{\alpha\beta\mu\nu\lambda}(k_1, k_2;
Q)\,=\,\int\,\frac{d^4 k}{\pi^{2}\,i}\nonumber\\ 
\hspace{-0.3in}&&\times {\rm
tr}\,\Bigg\{\sigma^{\alpha\,\beta}\,\frac{1}{M_{\rm
N}-\hat{k}-\hat{Q}}\,\sigma^{\mu\,\nu}\,\frac{1}{M_{\rm
N}-\hat{k}-\hat{Q}-\hat{k}_1}\,\gamma^{\lambda}\,\frac{1}{M_{\rm
N}-\hat{k}-\hat{Q}-\hat{k}_1-\hat{k}_2}\Bigg\}.\nonumber\\
\hspace{-0.5in}&&
\end{eqnarray}
The 4--vector $Q = a\,k_1 + b\,k_2$ is an arbitrary shift of a virtual
momentum, where $a$ and $b$ are arbitrary parameters. The
$Q$--dependent part of the structure function we obtain by using the
Gertsein--Jackiw method [15]
\begin{eqnarray}\label{label4.32}
\delta\,{\cal J}^{\alpha\beta\mu\nu\lambda}(k_1, k_2;
Q)&=&-\,\frac{1}{12}\,{\rm
tr}(\gamma_{\rho}\,\sigma^{\alpha\,\beta}\,\gamma^{\rho}\,
\sigma^{\mu\,\nu}\,\hat{Q}\,\gamma^{\lambda}\,+\,
\sigma^{\alpha\,\beta}\,\gamma_{\rho}\,
\sigma^{\mu\,\nu}\,\gamma^{\rho}\,\hat{Q}\,
\gamma^{\lambda}\,+\nonumber\\ &&
+\,\sigma^{\alpha\,\beta}\,\hat{Q}\,\sigma^{\mu\,\nu}\,
\gamma_{\rho}\,\gamma^{\lambda}\,\gamma^{\rho})= \frac{1}{6}\,{\rm
tr}(\sigma^{\alpha\,\beta}\,\hat{Q}\,\sigma^{\mu\,\nu}\,\gamma^{\lambda}\,).
\end{eqnarray}
Now we should proceed to the evaluation of ${\cal
J}^{\alpha\beta\mu\nu\lambda}(k_1, k_2; Q)$. By analogy
with ${\cal J}^{\beta\alpha\mu}(k_1, k_2; Q)$ we get
\begin{eqnarray}\label{label4.33}
\hspace{-0.3in}&&{\cal J}^{\alpha\beta\mu\nu\lambda}(k_1, k_2; Q)
\,=\nonumber\\
\hspace{-0.3in}&&=\,\int\,\frac{d^4 k}{\pi^{2}\,i}\,
{\rm tr}\Bigg\{\sigma^{\alpha\,\beta}\,
\frac{M_{\rm N}+\hat{k}+\hat{Q}}{M^{2}_{\rm N}-k^{2}}\,
\Bigg[1\,+\,\frac{2\,k\cdot Q}{M^{2}_{\rm N}-k^{2}}\Bigg]\,
\sigma^{\mu\,\nu}\,
\frac{M_{\rm N}+\hat{k}+\hat{Q}+\hat{k}_1}{M^{2}_{\rm N}-k^{2}}
\nonumber\\
\hspace{-0.3in}&&\times\,
\Bigg[1\,+\,\frac{2\,k\cdot (Q+k_1)}{M^{2}_{\rm N}-k^{2}}\Bigg]
\,\gamma^{\lambda}\,
\frac{M_{\rm N}+\hat{k}+\hat{Q}+\hat{k}_1+
\hat{k}_2}{M^{2}_{\rm N}-k^{2}}\nonumber\\
\hspace{-0.3in}&&\times\,\Bigg[1\,+\,\frac{2\,k\cdot
(Q+k_1+k_2)}{M^{2}_{\rm N}- k^{2}}\Bigg]\Bigg\}\,=\nonumber\\
\hspace{-0.3in}&&=\,\int\frac{d^4 k}{\pi^2 i}\,\frac{1}{(M^{2}_{\rm
N}-k^{2})^{3}}{\rm tr}\,\{M^{2}_{\rm
N}[\sigma^{\alpha\,\beta}(\hat{k}+\hat{Q})\sigma^{\mu\,\nu}
\gamma^{\lambda}+\sigma^{\alpha\,\beta}\sigma^{\mu\,\nu}
(\hat{k}+\hat{Q}+\hat{k}_1)\gamma^{\lambda}\nonumber\\
\hspace{-0.5in}&&+\sigma^{\alpha\,\beta}\sigma^{\mu\,\nu}\gamma^{\lambda}
(\hat{k}+\hat{Q}+\hat{k}_1+\hat{k}_2)]+\sigma^{\alpha\,\beta}
(\hat{k}+\hat{Q})\sigma^{\mu\,\nu}(\hat{k}+\hat{Q}+\hat{k}_1)
\gamma^{\lambda}\nonumber\\
\hspace{-0.3in}&&\times (\hat{k}+\hat{Q}+\hat{k}_1+\hat{k}_2)\}\,
\Bigg[1\,+\,\frac{2\,k\cdot (3Q+2k_1+k_2)}{M^{2}_{\rm
N}-k^{2}}\Bigg]\,=\nonumber\\ 
\hspace{-0.3in}&&=\,\int\frac{d^4 k}{\pi^2
i}\frac{1}{(M^{2}_{\rm N}-k^{2})^{3}} {\rm tr}\,\{M^{2}_{\rm
N}[\sigma^{\alpha\,\beta}\hat{Q}\sigma^{\mu\,\nu}
\gamma^{\lambda}+\sigma^{\alpha\,\beta}\sigma^{\mu\,\nu}
(\hat{Q}+\hat{k}_1)\gamma^{\lambda}\nonumber\\
\hspace{-0.3in}&&+\sigma^{\alpha\,\beta}\sigma^{\mu\,\nu}
\gamma^{\lambda}(\hat{Q}+\hat{k}_1+\hat{k}_2)]-
\frac{1}{2}k^{2}\sigma^{\alpha\,\beta}\hat{Q}
\sigma^{\mu\,\nu}\gamma^{\lambda}\}\nonumber\\
\hspace{-0.3in}&&+\,2\,\int\frac{d^4 k}{\pi^2 i}
\frac{1}{(M^{2}_{\rm N}-k^{2})^4}{\rm tr}\,
\{\frac{1}{2}M^{2}_{\rm N}k^{2}
[\sigma^{\alpha\,\beta}(3\,\hat{Q}+2\,
\hat{k}_1+\hat{k}_2)\sigma^{\mu\,\nu}\gamma^{\lambda}\nonumber\\
\hspace{-0.3in}&&+\sigma^{\alpha\,\beta}
\sigma^{\mu\,\nu}(3\,\hat{Q}+2\,\hat{k}_1+\hat{k}_2)
\gamma^{\lambda}+\sigma^{\alpha\,\beta}\sigma^{\mu\,\nu}
\gamma^{\lambda}(3\,\hat{Q}+2\,\hat{k}_1+\hat{k}_2)]\nonumber\\
\hspace{-0.3in}&&-\frac{1}{6}\,k^4
\sigma^{\alpha\,\beta}(3\,\hat{Q}+2\,\hat{k}_1+\hat{k}_2)
\sigma^{\mu\,\nu}\gamma^{\lambda}\}\,=\,\nonumber\\
\hspace{-0.3in}&&={\cal J}^{\alpha\beta\mu\nu\lambda}_{(1)}(k_1, k_2;
Q)\,+\,{\cal J}^{\alpha\beta\mu\nu\lambda}_{(2)}(k_1, k_2; Q).
\end{eqnarray}
Integrating over $k$ we obtain
\begin{eqnarray}\label{label4.34}
&&{\cal J}^{\alpha\beta\mu\nu\lambda}_{(1)}(k_1, k_2;
Q)\,=\,\frac{1}{4}\,[1+2\,J_2(M_{\rm N})]\,{\rm
tr}(\sigma^{\alpha\,\beta}\hat{Q}\sigma^{\mu\,\nu}\gamma^{\lambda})\,
\nonumber\\
&&+\frac{1}{2}\,{\rm
tr}\,[\sigma^{\alpha\,\beta}\sigma^{\mu\,\nu}(\hat{Q}+\hat{k}_1)
\gamma^{\lambda}+\sigma^{\alpha\,\beta}\sigma^{\mu\,\nu}
\gamma^{\lambda}(\hat{Q}+\hat{k}_1+\hat{k}_2)],\nonumber\\
&&{\cal J}^{\alpha\beta\mu\nu\lambda}_{(2)}(k_1, k_2; Q)\,=
\,-\frac{1}{6}\,[-\frac{5}{6}+J_2(M_{\rm N})]\,
{\rm tr}[\sigma^{\alpha\,\beta}(3\hat{Q}+2\hat{k}_1+\hat{k}_2)
\sigma^{\mu\,\nu}\gamma^{\lambda}]\nonumber\\
&&-\frac{1}{6}\,{\rm tr}\,[\sigma^{\alpha\,\beta}
(3\hat{Q}+2\hat{k}_1+\hat{k}_2)\sigma^{\mu\,\nu}\gamma^{\lambda}\nonumber\\
&&+\sigma^{\alpha\,\beta}\sigma^{\mu\,\nu}(3\hat{Q}+2\hat{k}_1+\hat{k}_2)
\gamma^{\lambda}+\sigma^{\alpha\,\beta}
\sigma^{\mu\nu}\gamma^{\lambda}(3\hat{Q}+2\hat{k}_1+\hat{k}_2)].
\end{eqnarray}
Now we should sum up the contributions and collect like terms
\begin{eqnarray}\label{label4.35}
&&{\cal J}^{\alpha\beta\mu\nu\lambda}(k_1, k_2;
Q)=\frac{1}{6}\,{\rm
tr}[\sigma^{\alpha\,\beta}(\hat{Q}-\frac{1}{6}(2\,\hat{k}_1+
\hat{k}_2))\sigma^{\mu\,\nu}\gamma^{\lambda}]\nonumber\\
&&+ \frac{1}{6}\,{\rm tr}\,[
\sigma^{\alpha\,\beta}\sigma^{\mu\,\nu}(\hat{k}_1-
\hat{k}_2)\gamma^{\lambda}]\,+\,\frac{1}{6}\,{\rm
tr}\,[\sigma^{\alpha\,\beta}\sigma^{\mu\,\nu}
\gamma^{\lambda}(\hat{k}_1\,+\,2\,\hat{k}_2)].
\end{eqnarray}
It is seen that the $Q$--dependence agrees with that obtained by
the Gertsein--Jackiw method. Due to arbitrariness of $Q$ the
vector $(2\,\hat{k}_1+\hat{k}_2)/6$ can be removed by the redefinition of
$Q$. This gives
\begin{eqnarray}\label{label4.36}
&&{\cal J}^{\alpha\beta\mu\nu\lambda}(k_1, k_2; Q) = \frac{1}{6}\,{\rm
tr}(\sigma^{\alpha\beta} \hat{Q} \sigma^{\mu\nu}
\gamma^{\lambda})\nonumber\\ 
&&+ \frac{1}{6}\,{\rm tr}\,[\sigma^{\alpha\beta}
\sigma^{\mu\nu}(\hat{k}_1-\hat{k}_2)\gamma^{\lambda}] + \frac{1}{6}\,{\rm
tr}\,
[\sigma^{\alpha\beta}\sigma^{\mu\,\nu}\gamma^{\lambda}
(\hat{k}_1\,+\,2\,\hat{k}_2)].
\end{eqnarray}
By evaluating the traces over Dirac matrices we obtain the structure
function leading to the following effective Lagrangian [15]
\begin{eqnarray}\label{label4.37}
\hspace{-0.3in}&&{\cal L}_{\rm Fig.2d}(x)\,=
\,(-\,i\,e)\,\frac{g^{2}_{\rm T}}{4\,\pi^{2}}\,
\frac{1}{M^{2}_{\rm D}}\,\Big[\frac{8}{3}\,a\,
\partial_{\lambda}\,D^{\dagger\,\lambda\,\nu}(x)\,
D_{\nu\,\mu}(x)\,A^{\mu}(x)\nonumber\\
\hspace{-0.3in}&&+\,\frac{8}{3}\,a\,
D^{\dagger\,\mu\,\nu}(x)\,\partial_{\lambda}\,
D^{\lambda\,\nu}\,A_{\mu}(x)\,+\,\frac{8}{3}\,(b\,+\,a)\,
D^{\dagger}_{\mu\,\nu}(x)\,D^{\nu\,\lambda}(x)\,
\partial^{\mu}\,A_{\lambda}(x)\nonumber\\
\hspace{-0.3in}&&+\,\frac{8}{3}\,(b - a)\,
D^{\dagger}_{\mu\,\nu}(x)\,D^{\nu\,\lambda}(x)\,
\partial_{\lambda}\,A^{\mu}(x)\,-\,\frac{16}{3}\,D^{\dagger}_{\mu\,\nu}(x)\,
D^{\nu\,\lambda}(x)\,\partial^{\mu}\,A_{\lambda}(x)\,\nonumber\\
\hspace{-0.3in}&& +\,8\,
D^{\dagger}_{\mu\,\nu}(x)\,D^{\nu\,\lambda}(x)\,
(\partial^{\mu}\,A_{\lambda}(x)
- \partial_{\lambda}\,A^{\mu}(x))\Big].
\end{eqnarray}
For the derivation of the effective Lagrangian (\ref{label4.37}) we have
used the equation of motion
\begin{eqnarray}\label{label4.38}
\partial_{\lambda}\,D_{\mu\nu}(x)\,+\,\partial_{\mu}\,D_{\nu\lambda}(x)\,+\,\partial_{\nu}\,D_{\lambda\mu}(x)\,=\,0.
\end{eqnarray}
The analogous equation of motion is valid for the conjugated
field. The term proportional to $k^{\lambda}_2$ contributing to the
effective Lagrangian in the form of a divergence of the vector
potential of the electromagnetic field
$\partial^{\lambda}\,A_{\lambda}(x)$ can be omitted by singling
out the Lorentz gauge constraint for the electromagnetic potential,
i.e. $\partial^{\lambda}\,A_{\lambda}(x)\,=\,0$.

Collecting like terms in (\ref{label4.37}) we get
\begin{eqnarray}\label{label4.39}
&&{\cal L}_{\rm Fig.2d}(x)\,=\,(-\,i\,e)\,\frac{g^{2}_{\rm
T}}{4\,\pi^{2}}\,\frac{1}{M^{2}_{\rm D}}\,\Big[\frac{8}{3}\,(b\,+\,a -
1)\,D^{\dagger}_{\mu\nu}(x)\,D^{\nu\lambda}(x)\,
\partial^{\mu}\,A_{\lambda}(x)\nonumber\\
&&+\,\frac{8}{3}\,(b - a -
3)\,D^{\dagger}_{\mu\nu}(x)\,D^{\nu\lambda}(x)\,
\partial_{\lambda}\,A^{\mu}(x)+\frac{8}{3}\,a\,
\partial_{\lambda}\,D^{\dagger\lambda\,\nu}(x)\,D_{\nu\mu}(x)\,
A^{\mu}(x)\nonumber\\
&&+\,\frac{8}{3}\,a\,D^{\dagger \mu\nu}(x)\,\partial_{\lambda}\,
D^{\lambda\nu}(x)\,A_{\mu}(x)\Big].
\end{eqnarray}
The third and the fourth terms can be reduced by applying the equation
of motion
\[\partial_{\lambda}\,D^{\lambda\,\nu}(x)\,= - M^{2}_{\rm D}\,D^{\nu}(x)\,\]
and analogous for the conjugated field. Then, setting
\begin{eqnarray}\label{label4.40}
b\,+\,a - 1\,= - b\,+\,a\,+\,3
\end{eqnarray}
we obtain $b\,=\,2$ and bring up the effective Lagrangian
(\ref{label4.39}) to the following irreducible form
\begin{eqnarray}\label{label4.41}
\hspace{-0.3in}&&{\cal L}_{\rm Fig.2d}(x)\,=\nonumber\\
\hspace{-0.3in}&&=\,i\,e\,\frac{2\,g^{2}_{\rm
T}}{3\,\pi^{2}}\,a\,[-\,D^{\dagger}_{\mu\nu}(x)\,
A^{\nu}(x)\,D^{\mu}(x) +  D^{\mu\nu}(x)\,A_{\nu}(x)\,D
^{\dagger}_{\mu}(x)]\,\nonumber\\ 
\hspace{-0.3in}&&
+\,i\,e\,\frac{2\,g^{2}_{\rm T}}{3\,\pi^{2}}\,\frac{1}{M^{2}_{\rm
D}}\,(1\,+\,a)\,D^{\dagger}_{\mu\nu}(x)\,
D^{\nu\,\lambda}(x)\,(\partial_{\lambda}\,
A^{\mu}(x) - \partial^{\mu}\,A_{\lambda}(x)).
\end{eqnarray}
The first two terms define the finite contributions to the
renormalization constant of the wave function of the deuteron, whereas
the last term coincides with the well--known phenomenological
interaction ${\cal L}_{\rm A}(x)$ introduced by Aronson [19]
\begin{eqnarray}\label{label4.42}
{\cal L}_{\rm A}(x) = i\,e\,\frac{2\,g^{2}_{\rm T}}{3\,\pi^{2}}\,
\frac{1}{M^{2}_{\rm
D}}\,(1 + a)\,D^{\dagger}_{\mu\nu}(x)\,
D^{\nu\lambda}(x)\,{F_{\lambda}}^{\mu}(x)
\end{eqnarray}
for the description of the charged vector field coupled to an
electromagnetic field.

Thus, we have shown that the anomaly of the one--nucleon loop triangle
$VTT$--diagram, where $V$ and $T$ stand for the vector and tensor
vertices determined by the Dirac matrices $\gamma^{\alpha}$ and
$\sigma_{\mu\nu}$, respectively, calculated at leading order in the
large $N_C$ expansion defines fully the phenomenological Aronson
Lagrangian describing the deuteron coupled to an external
electromagnetic field.

\subsection{The magnetic dipole and electric quadrupole moments of 
the deuteron}

\hspace{0.2in} The effective Lagrangian describing both the magnetic
dipole and electric quadrupole moments of the deuteron is determined
by the sum of ${\cal L}_{\rm Fig.2c}(x)$ and ${\cal L}_{\rm
Fig.2d}(x)$ given by Eq.(\ref{label4.28}) and Eq.(\ref{label4.41}),
respectively, and reads
\begin{eqnarray}\label{label4.43}
\delta {\cal L}^{\rm
el}_{\rm eff}(x)&=&i\,e\,\frac{bg^{2}_{\rm
V}-4ag^{2}_{\rm T}}{6\,\pi^{2}}\,D^{\dagger}_{\mu\nu}(x)\,A^{\nu}(x)\,D^{\mu}(x)\nonumber\\
&-&i\,e\,\frac{bg^{2}_{\rm
V} - 4ag^{2}_{\rm T}}{6\,\pi^{2}}\,D^{\mu\nu}(x)\,A_{\nu}(x)\,D^{\dagger}_{\mu}(x)\nonumber\\
&+&i\,e\,\frac{g^{2}_{\rm
V}}{6\,\pi^{2}}\,(-\,2\,b\,+\,3)\,D^{\dagger}_{\mu}(x)\,
D_{\nu}(x)\,F^{\mu\nu}(x)\nonumber\\
&+&i\,e\,(1\,+\,a)\,\frac{2\,g^{2}_{\rm
T}}{3\,\pi^{2}}\,\frac{1}{M^{2}_{\rm
D}}\,D^{\dagger}_{\mu\nu}(x)\,D^{\nu\lambda}(x)
\,{F_{\lambda}}^{\mu}(x),
\end{eqnarray}
where $a$ and $b$ are arbitrary parameters related to the ambiguities
of the one--nucleon loop diagrams with respect to a shift of a virtual
nucleon momentum. We consider them as free parameters of the approach
[15].

In order to fix these parameters it is convenient to write down the
total effective Lagrangian of the physical deuteron coupled to an
external electromagnetic field
\begin{eqnarray}\label{label4.44}
{\cal L}^{\rm
el}_{\rm eff}(x)&=&
-\frac{1}{2}\,D^{\dagger}_{\mu\nu}(x)D^{\mu\nu}(x) + M^2_{\rm
D}\,D^{\dagger}_{\mu}(x)D^{\mu}(x)\nonumber\\
&+&i\,e\,\frac{bg^{2}_{\rm
V}-4ag^{2}_{\rm T}}{6\,\pi^{2}}\,D^{\dagger}_{\mu\nu}(x)\,A^{\nu}(x)\,D^{\mu}(x)\nonumber\\
&-&i\,e\,\frac{bg^{2}_{\rm
V} - 4ag^{2}_{\rm T}}{6\,\pi^{2}}\,D^{\mu\nu}(x)\,A_{\nu}(x)\,D^{\dagger}_{\mu}(x)\nonumber\\
&+&i\,e\,\frac{g^{2}_{\rm
V}}{6 \pi^{2}}\,(-\,2\,b\,+\,3)\,D^{\dagger}_{\mu}(x)\,
D_{\nu}(x)\,F^{\mu\nu}(x)\nonumber\\
&+&i\,e\,(1\,+\,a)\,\frac{2 g^{2}_{\rm
T}}{3 \pi^{2}}\,\frac{1}{M^{2}_{\rm
D}}\,D^{\dagger}_{\mu\nu}(x)\,D^{\nu\lambda}(x)
\,{F_{\lambda}}^{\mu}(x).
\end{eqnarray}
Two terms having the structure
$D^{\mu\nu}(x)A_{\mu}(x)D^{\dagger}_{\nu}(x)$ and
$D^{\dagger}_{\mu\nu}(x)A^{\mu}(x)D^{\nu}(x)$ should describe the
interaction of the deuteron with an external electromagnetic field
included by a minimal way, whilst the last two terms are responsible
for the non--trivial contributions to the magnetic dipole and electric
quadrupole moments of the deuteron.  In terms of the parameters of the
effective interactions Eq.(\ref{label4.44}) the magnetic dipole moment
$\mu_{\rm D}$, measured in nuclear magnetons, and the electric
quadrupole moment $Q_{\rm D}$, measure in ${\rm fm}^2$, of the
deuteron are given by
\begin{eqnarray}\label{label4.45}
\mu_{\rm D} &=&(1\,+\,a)\frac{g^{2}_{\rm T}}{3\pi^{2}} +
(3\,-\,2\,b)\,\frac{g^{2}_{\rm V}}{12\pi^{2}},\nonumber\\ 
Q_{\rm D} &=& \Bigg[(2\,+\,2\,a)\,\frac{g^{2}_{\rm T}}{3\pi^{2}} -
(3\,-\,2\,b)\,\frac{g^{2}_{\rm V}}{6\pi^{2}}\Bigg]\,\frac{1}{M^2_{\rm
D}}
\end{eqnarray}
at the constraint 
\begin{eqnarray}\label{label4.46}
b\,\frac{g^{2}_{\rm V}}{6\pi^{2}}\,-\,2\,a\,\frac{g^{2}_{\rm
 T}}{3\pi^{2}} = 1
\end{eqnarray}
reducing the first two terms in effective Lagrangian
Eq.(\ref{label4.43}) to the standard minimal form which can be
obtained from the effective Lagrangian of the free deuteron field by
the shift $\partial_{\mu}D_{\nu}(x) \to (\partial_{\mu} +
\,i\,e\,A_{\mu}(x)) D_{\nu}(x)$.

Retaining the former relation between the electric quadrupole moment
and the coupling constant $g_{\rm V}$, $Q_{\rm D} = 2g^2_{\rm V}/\pi^2
M^2_{\rm D}$ [15] that gives $g_{\rm V} = 11.319$, we express the
parameters $a$ and $b$ in terms of the coupling constants $g_{\rm V}$,
$g_{\rm T}$ and the magnetic dipole moment $\mu_{\rm D}$:
\begin{eqnarray}\label{label4.47}
a = -1 + \frac{3}{2}\,\frac{\pi^2}{g^2_{\rm T}}\Bigg(\mu_{\rm D} +
\frac{g^2_{\rm V}}{\pi^2}\Bigg)\quad,\quad b = \frac{9}{2} -
\frac{3\pi^2}{g^2_{\rm V}}\,\mu_{\rm D}.
\end{eqnarray}
Substituting Eq.(\ref{label4.47}) in Eq.(\ref{label4.46}) we get the
relation between coupling constants $g_{\rm V}$, $g_{\rm T}$ and the
magnetic dipole moment $\mu_{\rm D}$
\begin{eqnarray}\label{label4.48}
g_{\rm T} = \sqrt{\frac{3}{8}\,g^2_{\rm V} +
\frac{3}{2}\,\pi^2\Bigg(1 + \frac{3}{2}\,\mu_{\rm
D}\Bigg)}=0.799\,g_{\rm V}.
\end{eqnarray}
The numerical value $g_{\rm T} = 0.799\,g_{\rm V}$ we obtain at
$g_{\rm V} = 11.319$ [15], $\mu_{\rm D}=0.857$ [22] and $N_C = 3$. The
sign of the coupling constant $g_{\rm T}$ should coincide with the
sign of the coupling constant $g_{\rm V}$. For the opposite sign the
binding energy $\varepsilon_{\rm D}$ given by Eq.(\ref{label4.8})
becomes negative that means the absence of the bound neutron--proton
state with quantum numbers of the deuteron.

For the evaluation of the binding energy $\varepsilon_{\rm D}$
determined by Eq.(\ref{label4.8}) we should keep only the leading
contribution to the coupling constant $g_{\rm T}$ in the large $N_C$
expansion, i.e.
\begin{eqnarray}\label{label4.49}
g_{\rm T} = \sqrt{\frac{3}{8}}\,g_{\rm V} + O(1/\sqrt{N_C}).
\end{eqnarray}
Substituting this relation into Eq.(\ref{label4.8}) we can describe
the experimental value of the binding energy of the deuteron
$\varepsilon_{\rm D} = 2.225\,{\rm MeV}$ at the cut--off $\Lambda_{\rm
D} = 46.172\,{\rm MeV}$. The spatial region of virtual nucleon field
fluctuations forming the physical deuteron related to this value of
the cut--off $1/\Lambda_{\rm D} \sim r_{\rm D} = 4.274\,{\rm fm}$
agrees good with the experimental value of the radius of the deuteron
$r_{\rm D} = (4.31895\pm 0.00009)\,{\rm fm}$ [22]. This result
confirms estimates obtained in Ref.\,[15].

The effective Lagrangian of the deuteron field coupled to an external
electromagnetic field is given by
\begin{eqnarray}\label{label4.50}
\hspace{-0.5in}&&{\cal L}^{\rm
el}_{\rm eff}(x) =
-\frac{1}{2}\,[(\partial_{\mu} -
\,i\,e\,A_{\mu}(x))D^{\dagger}_{\nu}(x) - (\partial_{\nu} -
\,i\,e\,A_{\nu}(x))D^{\dagger}_{\mu}(x)]\nonumber\\
\hspace{-0.5in}&&\times\,[(\partial^{\mu} + \,i\,e\,A^{\mu}(x))D^{\nu}(x) -
(\partial^{\nu} + \,i\,e\,A^{\nu}(x))D^{\mu}(x)] + M^2_{\rm
D}\,D^{\dagger}_{\mu}(x)D^{\mu}(x)\nonumber\\ 
\hspace{-0.5in}&&+\,i\,e\,\Bigg(\mu_{\rm D}\, - \,\frac{1}{2}\,Q_{\rm
D}\,M^2_{\rm D}\Bigg)\,D^{\dagger}_{\mu}(x)\,
D_{\nu}(x)\,F^{\mu\nu}(x)\nonumber\\
\hspace{-0.5in}&&+\,i\,e\,\Bigg(\mu_{\rm D}\, + \,\frac{1}{2}\,Q_{\rm
D}\,M^2_{\rm D}\Bigg)\frac{1}{M^2_{\rm D}}\,D^{\dagger}_{\mu\nu}(x)
\,D^{\nu\lambda}(x)
\,{F_{\lambda}}^{\mu}(x).
\end{eqnarray}
The term of order $O(e^2)$ can be also derived in the NNJL model by
using shift ambiguities of one--nucleon loop diagrams. This term is
required by the electromagnetic gauge invariance of the effective
Lagrangian of the deuteron field coupled to an external
electromagnetic field, but it does not affect on the electromagnetic
parameters of the deuteron which are of order $O(e)$.

\section{Conclusion}

\hspace{0.2in} We have shown that the Nambu--Jona--Lasinio model of
light nuclei or the NNJL model as well as the ENJL model with chiral
$U(3)\times U(3)$ symmetry [4--11] is motivated by QCD. The NNJL model
describes low--energy nuclear forces in the nuclear phase of QCD in
terms of one--nucleon loop exchanges. One--nucleon loop exchanges
provide a minimal way of the transfer of nucleon flavours from an
initial to a final nuclear state and allow to take into account
contributions of nucleon--loop anomalies. These anomalies are related
to high--energy fluctuations of virtual nucleon fields, i.e. the
$N\bar{N}$ fluctuations, and fully determined by one--nucleon loop
diagrams [29--31]. The dominance of contributions of one--nucleon loop
anomalies to effective Lagrangians describing low--energy interactions
of the deuteron coupled to itself, nucleons and other particles we
justify within the large $N_C$ expansion in QCD with $SU(N_C)$ gauge
group at $N_C \to \infty$. It is well--known that anomalies of
quark--loop diagrams play an important role for the correct
description of strong low--energy interactions of low--lying hadrons
[4--11].  We argue an important role of nucleon--loop anomalies for
the correct description of low--energy nuclear forces in the nuclear
physics.

It should be emphasized that nucleon--loop anomalies can be
interpreted as non--trivial contributions of the non--perturbative
quantum vacuum -- the nucleon Dirac sea [32].  In nuclear physics the
influence of the nucleon Dirac sea on low--energy properties of finite
nuclei has been analysed within quantum field theoretic approaches in
the one--nucleon loop approximation [33].  Unfortunately, in these
approaches contributions of one--nucleon loop anomalies have not been
taken into account. The NNJL model allows to fill this blank.

For the derivation of the NNJL model from the first principles of QCD we
distinguish three non--perturbative phases of QCD: 1) the
low--energy quark--gluon phase (low--energy QCD), 2) the hadronic
phase and 3) the nuclear phase. Skipping over the intermediate
low--energy quark--gluon phase by means of the integration over high--
and low--energy quark and gluon fluctuations one arrives at the
hadronic phase of QCD containing only local hadron fields with quantum
numbers of mesons and baryons coupled at energies below the SB$\chi$S
scale $\Lambda_{\chi} \simeq 1\,{\rm GeV}$. The couplings of
low--lying mesons with masses less than the SB$\chi$S scale to
low--lying octet and decuplet of baryons can be described by Effective
Chiral Lagrangians with chiral $U(3)\times U(3)$ symmetry.

Integrating in the hadronic phase of QCD over heavy hadron degrees of
freedom with masses exceeding the SB$\chi$S scale one arrives at the
nuclear phase of QCD which characterizes itself by the appearance of
bound nucleon states -- nuclei. At low energies the result of
integration over heavy hadron degrees of freedom can be represented in
the form of phenomenological local many--nucleon interactions. Some of
these interactions are responsible for creation of many--nucleon
collective excitations which acquire the properties of observed nuclei
through nucleon--loop and low--lying meson exchanges.  This effective
field theory describes nuclei and processes of their low--energy
interactions by considering nuclei as elementary particles represented
by local interpolating fields.

Following this scenario of the description of nuclei and their
low--energy interactions from the first principles of QCD the deuteron
should be produced in the nuclear phase of QCD by a phenomenological
local four--nucleon interaction as the Cooper np--pair with quantum
numbers of the deuteron. The low--energy parameters of the physical
deuteron, i.e. the binding energy, the magnetic dipole $\mu_{\rm D}$
and electric quadrupole $Q_{\rm D}$ moments and so, the Cooper
np--pair acquires through one--nucleon loop exchanges. We have shown
that the main part of the kinetic term of the effective Lagrangian of
the free physical deuteron field is induced by the contribution of
high--energy (short--distance) fluctuations of virtual nucleon fields
related to the anomaly of the one--nucleon loop $VV$--diagram with two
vector vertices.

In turn, the magnetic dipole $\mu_{\rm D}$ and electric quadrupole
$Q_{\rm D}$ moments of the physical deuteron are fully determined by
high--energy (short--distance) fluctuations of virtual nucleon fields
related to the anomalies of the triangle one--nucleon loop $VVV$-- and
$VTT$--diagrams.  Thus, high--energy (short--distance) fluctuations of
virtual nucleon fields related to anomalies of one--nucleon loop
diagrams play a dominant role for the correct description of
electromagnetic properties of the physical deuteron in the NNJL model.

As regards low--energy (long--distance) fluctuations of virtual
nucleon fields they give a significant contribution only to the
binding energy of the deuteron $\varepsilon_{\rm D}$. The strength of
low--energy (long--distance) fluctuations of virtual nucleon fields is
restricted by the cut--off $\Lambda_{\rm D} = 46.172\,{\rm MeV}$. The
spatial region of virtual nucleon field fluctuations forming the
physical deuteron related to this value of the cut--off
$1/\Lambda_{\rm D} \sim r_{\rm D} = 4.274\,{\rm fm}$ agrees good with
the experimental value of the radius of the deuteron $r_{\rm D} =
(4.31895\pm 0.00009)\,{\rm fm}$ [22]. This confirms our estimates
obtained in Ref.\,[15].

It is well--known that in the potential model approach to the
description of the deuteron the electric quadrupole moment of the
deuteron $Q_{\rm D}$ is caused by nuclear tensor forces which are of
great deal of importance for the existence of the deuteron as a bound
np--state [34].

The proportionality of the coupling constant of the phenomenological
local four--nucleon interaction Eq.(\ref{label3.3}), responsible for
creation of the Cooper np--pair with quantum numbers of the deuteron,
and the binding energy of the deuteron $\varepsilon_{\rm D}$
Eq.(\ref{label3.20}) to the electric quadrupole moment $Q_{\rm D}$
testifies an important role of nuclear tensor forces for the formation
of the deuteron in the NNJL model.

To the evaluation of one--nucleon loop diagrams defining effective
Lagrangians describing processes of low--energy interactions of the
deuteron coupled to itself and an electromagnetic field we apply
expansions in powers of the momenta of interacting particles and keep
only leading terms of the expansions. This approximation can be
justified in the large $N_C$ expansion. Indeed, in QCD with the
$SU(N_C)$ gauge group at $N_C \to \infty$ the nucleon mass is
proportional to the number of quark colours [17]: $M_{\rm N} \sim
N_C$.  Since for the derivation of effective Lagrangians describing
the deuteron and amplitudes of low--energy nuclear processes all
external momenta of interacting particles should be kept off--mass
shell, the masses of virtual nucleon fields are larger compared with
the external momenta. An expansion of one--nucleon loop diagrams in
powers of $1/M_{\rm N}$ giving an external momentum expansion
corresponds to the expansion in powers of $1/N_C$. In this case the
leading order in the large $N_C$ expansion gives the leading order
contributions in the expansion in powers of external momenta of
interacting particles.  We should emphasize that anomalous
contributions of one--nucleon loop diagrams are determined by the
least powers in external momentum expansions. Thereby, the dominance
of contributions of nucleon--loop anomalies to effective Lagrangians
describing low--energy nuclear forces in the NNJL model is fully
supported by the large $N_C$ expansion. The accuracy of this
approximation is rather high. Indeed, the real parameter of the
expansion of one--nucleon loop diagrams is $1/M^2_{\rm N} \sim
1/N^2_C$ but not $1/M_{\rm N} \sim 1/N_C$. Thereby, next--to--leading
corrections should be of order $O(1/N^2_C)$.

The inclusion of the interaction of the deuteron field with the tensor
nucleon current Eq.(\ref{label4.1}) has given a possibility of the
self-consistent description of the electromagnetic properties of the
deuteron, the magnetic dipole moment $\mu_{\rm D}$ and the electric
quadrupole moment $Q_{\rm D}$, in terms of effective interactions of
the Corben--Schwinger and Aronson kinds induced by one--nucleon loop
diagrams. By fitting the experimental values of the magnetic dipole
moment $\mu_{\rm D} = 0.857$, measured in nucleon magnetons $\mu_{\rm
N} = e/2M_{\rm N}$, and the electric quadrupole moment $Q_{\rm D} =
0.286$, measured in ${\rm fm}^2$, supplemented by the requirement of
the electromagnetic gauge invariance of the effective Lagrangian of
the deuteron field coupled to an external electromagnetic field we
have got the relation between the coupling constants $g_{\rm V}$ and
$g_{\rm T}$: $g_{\rm T} = 0.799\,g_{\rm V}$ calculated at $N_C =
3$. At leading order in the large $N_C$ expansion we get $g_{\rm T} =
\sqrt{3/8}\,g_{\rm V} + O(1/\sqrt{N_C})$. This relation agrees good
with that obtained in Ref.\,[14] (see Eq.(16) of Ref.\,[14]). Due to
this relation the experimental value of the binding energy of the
deuteron can be described by the cut--off $\Lambda_{\rm D} =
46.172\,{\rm MeV}$. This corresponds to the spatial region of virtual
nucleon field fluctuations forming the physical deuteron
$1/\Lambda_{\rm D} \sim r_{\rm D} = 4.274\,{\rm fm}$ agreeing good
with the experimental value of the radius of the deuteron $r_{\rm D} =
(4.31895\pm 0.00009)\,{\rm fm}$ [22].  

For further applications of the NNJL model to the description of
low--energy nuclear reactions of astrophysical interest we anticipate
the results in agreement with those obtained in Refs.\,[35,36]. 

The quantum field theoretic scenario to treating nuclei as
many--nucleon collective excitations induced by phenomenological local
many--nucleon interactions allows a plain extension of the NNJL model
by the inclusion of light nuclei ${^3}{\rm He}$, ${^3}{\rm H}$ and
${^4}{\rm He}$ as three-- and four--nucleon collective
excitations. The binding energies and other low--energy parameters of
these excitations should be determined through nucleon--loop and
low--lying meson exchanges. 

On this way it is important to notice that the spinorial structure of
the operators of three--nucleon densities coupled to the ${^3}{\rm
He}$ and the ${^3}{\rm H}$ is very much restricted. One can show that
only the three--nucleon densities $[\bar{p^c}(x)\gamma^{\mu}\gamma^5
p(x)]\gamma_{\mu} n(x)$ and $[\bar{n^c}(x)\gamma^{\mu}\gamma^5
n(x)]\gamma_{\mu} p(x)$ can lead to the appearance of the bound
${^3}{\rm He}$ and ${^3}{\rm H}$ state, respectively. At the
quantum field theoretic level this result explains a well--known
experimental fact of the compensation of spins and magnetic dipole
moments of pp and nn pairs inside nuclei which has been put into the
foundation of the shell--model of nuclei [34].

The extension of the NNJL model by the inclusion of ${^3}{\rm He}$,
${^3}{\rm H}$ and ${^4}{\rm He}$ would give a possibility to analyse
within the NNJL model the reactions of the p--p chain [37] started with
the reaction p + p $\to$ D + e$^+$ + $\nu_{\rm e}$ and to apply the
extended version of the NNJL model to the description of the reactions
p + D $\to$ ${^3}{\rm He}$ + $\gamma$, p + ${^3}{\rm He}$ $\to$
${^4}{\rm He}$ + e$^+$ + $\nu_{\rm e}$ and so on.

Chiral perturbation theory can be naturally incorporated into the NNJL
model [35] in terms of Effective Chiral Lagrangians with chiral
$U(3)\times U(3)$ symmetry describing low--lying baryons and mesons
interacting at low energies [4--11].

The quantum field theoretic description of the deuteron within the
NNJL model can be also of use for the analysis of the properties of
dibaryons. Indeed, following Oakes [38] the deuteron can be considered
as a component of the $SU(3)_{\rm flavour}$ decuplet $\tilde{\bf
10}_f$ of dibaryons with $Y=2$ and $I=0$, where $Y$ and $I$ are the
hypercharge and the isotopical spin, respectively. In the chiral limit
the binding energies, the magnetic dipole and electric quadrupole
moments of dibaryons of the decuplet $\tilde{\bf 10}_f$ should be
equal. The splitting of the parameters of the components of the
decuplet $\tilde{\bf 10}_f$ can be obtained within Chiral perturbation
theory incorporated into the NNJL model.

\section*{Acknowledgement}

\hspace{0.2in} The authors (A.N. Ivanov and N.I. Troitskaya) are
grateful to Prof. Randjbar--Daemi, the Head of the High Energy Section
of the Abdus Salam International Centre for Theoretical Physics (ICTP)
in Trieste, for warm and kind hospitality extended to them
during the whole period of their stay at ICTP, when this work was
started.

\newpage

\section*{Figure caption}

\begin{itemize}
\item 

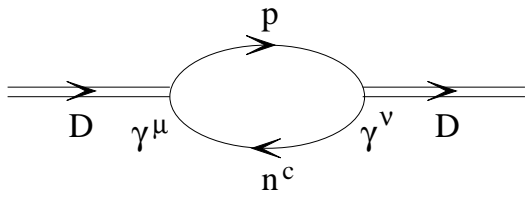
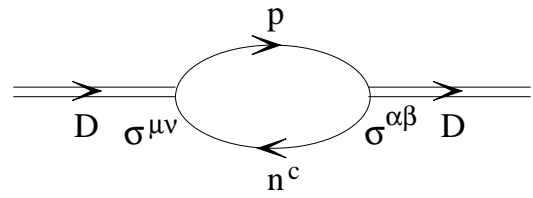
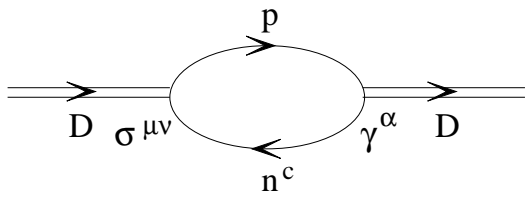
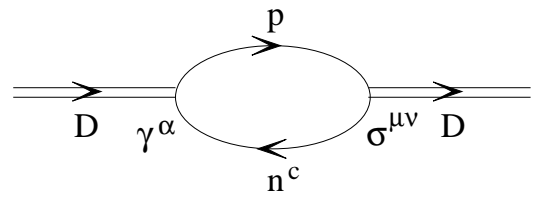
Fig.1. One--nucleon loop diagrams contributing in the NNJL model
to the binding energy of the physical deuteron, where $n^c =
C\,\bar{n}^T$ is the field of anti--nucleon.

\item 

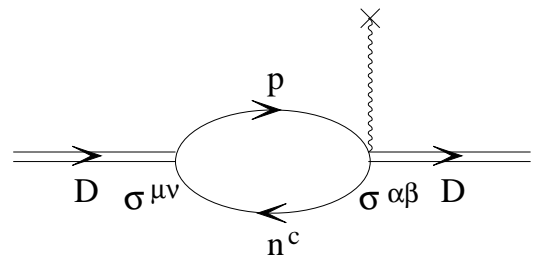
Fig.2. One--nucleon loop diagrams describing in the NNJL model the effective
Lagrangian of the deuteron coupled to an electromagnetic field through
the magnetic dipole and electric quadrupole moments, where $n^c =
C\,\bar{n}^T$ is the field of anti--nucleon.
\end{itemize}


\newpage

\begin{thebibliography}{99}
\bibitem{[1]} 
Y. Nambu and G. Jona--Lasinio, Phys. Rev. {\bf 122}, 345
(1961).
\bibitem{[2]}
Y. Nambu and G. Jona--Lasinio,
Phys. Rev. {\bf 124}, 246 (1961).
\bibitem{[3]}
J. Bardeen, L. N. Cooper and J. R. Schrieffer,
Phys. Rev. {\bf 106}, 162 (1957), ibid.{\bf 108}, 1175 (1957).  
\bibitem{[4]}
S. P. Klevansky,
Rev. Mod. Phys. {\bf 64}, (1992) 649 and references therein.
\bibitem{[5]} 
T. Hatsuda and T. Kunihiro, Phys. Rep. {\bf 247}, 221
(1994) and references therein.
\bibitem{[6]}
A. N. Ivanov, N. I. Troitskaya, M. Faber, M. Schaler and M. Nagy,
Nuovo Cim. {\bf A 107}, 1667 (1994); Phys. Lett. {\bf B 336}, 555 (1995);
A. N. Ivanov, N. I. Troitskaya and M. Faber,
Nuovo Cim. {\bf A 108}, 613 (1995).
\bibitem{[7]} 
A. N. Ivanov, M. Nagy and N. I. Troitskaya,
Int. J. Mod. Phys. {\bf A 7}, 7305 (1992); A. N. Ivanov,
Int. J. Mod. Phys. {\bf A 8}, 853 (1993); A. N. Ivanov,
N. I. Troitskaya and M. Nagy, Int. J. Mod. Phys. {\bf A 8}, 2027, 3425
(1993); Phys. Lett. {\bf B 295}, 308 (1992); Phys. Lett. {\bf B 308}, 111
(1993); A. N. Ivanov and N. I. Troitskaya, Nuovo Cimento {\bf A 108}, 555
(1995).
\bibitem{[8]}
A. N. Ivanov, M. Nagy and N. I. Troitskaya,
Phys. Rev. {\bf C 59}, 451 (1999);
Ya. A. Berdnikov, A. N. Ivanov, V. F. Kosmach and N. I. Troitskaya,
Phys. Rev. {\bf C 60},  015201 (1999).
\bibitem{[9]}
J. Bijnens, C. Bruno and E. de Rafael,
Nucl. Phys. {\bf B 390}, 501 (1993);
J. Bijnens, E. de Rafael and H. Zheng,
Z. Phys. {\bf C 62}, 437 (1994).
\bibitem{[10]} 
K. Kikkawa, Progr. Theor. Phys. {\bf 56}, (1976) 947;
H. Kleinert, {\it Proc. of Int. Summer School of Subnuclear Physics},
Erice 1976, Ed. A. Zichichi, p.289.  
\bibitem{[11]} 
A. Dhar, R. Shankar and S. R. Wadia,
Phys. Rev. {\bf D 31}, 3256 (1985);
D. Ebert and H. Reinhart,
Nucl. Phys. {\bf B 271}, 188 (1986);
M. Wakamatsu, 
Ann. of Phys. (N.Y.) {\bf 193}, 287 (1989).
\bibitem{[12]}
S. Gasiorowicz and D. A. Geffen,
Rev. Mod. Phys. {\bf 41}, 531 (1969) and references therein.
\bibitem{[13]}
J. Wess and B. Zumino,
Phys. Lett. {\bf B 37}, 95 (1971).
\bibitem{[14]} 
A. N. Ivanov, N. I. Troitskaya, M. Faber and
H. Oberhummer, 
Phys. Lett. {\bf B 361}, 74 (1995).
\bibitem{[15]} 
A. N. Ivanov, N. I. Troitskaya, M. Faber and
H. Oberhummer, Nucl. Phys. {\bf A 617}, 414 (1997) and Nucl.Phys. {\bf
A 625}, 896 (1997) (Erratum).
\bibitem{[16]}
G. 't Hooft,
Nucl. Phys. B75 (1974) 461.
\bibitem{[17]} 
E. Witten, 
Nucl. Phys. B160 (1979) 57.
\bibitem{[18]}
H. C. Corben and J. Schwinger,
Phys. Rev. {\bf 58}, 953 (1940).
\bibitem{[19]}
H. Aronson,
Phys. Rev. {\bf 186}, 1434 (1969).
\bibitem{[20]}
B. Sakita and C. J. Goebal,
Phys. Rev. {\bf 127}, 1787 (1962);
B. Sakita,
Phys. Rev. {\bf 127}, 1800 (1962).
\bibitem{[21]} 
C. W. Kim and H. Primakoff, 
{\it Nuclei as elementary
particles in weak and electromagnetic processes} in {\it MESONS IN
NUCLEI}, Vol.1 (1979) pp.67--106, ed. M. Rho and D. Wilkinson,
Noth--Holland Publishing Company Amsterdam--New York--Oxford;
C. W. Kim and H. Primakoff,
Phys. Rev. {\bf B 139}, 14447 (1965); ibid. {\bf B 140}, 586 (1965).
\bibitem{[22]} 
M. M. Nagels {\it et al.}, Nucl. Phys. {\bf B 147}, 253 (1979).
\bibitem{[23]}
A. V. Anisovich and V. A. Sadovnikova,
Europ. Phys. J. {\bf A 2}, 199 (1999).
\bibitem{[24]} 
V.V. Anisovich, M.N. Kobrinsky, D.I. Melikhov,
A.V. Sarantsev, Nucl. Phys. {\bf A 544}, 747 (1992); 
V.V. Anisovich ,
D.I. Melikhov, B.C. Metsch and H.R. Petry, Nucl. Phys. {\bf A 563},
549 (1993).
\bibitem{[25]}
M. K. Volkov and C. Wess,
Phys. Rev. {\bf D 56}, 221 (1997);
M.K. Volkov and V.L. Yudichev,
Int. J. Mod. Phys. {\bf A14}, 4621 (1999). 
\bibitem{[26]}
R. L. Jaffe, 
Phys. Rev. {\bf D 15}, 267, 281 (1977);
R. L. Jaffe and F. E. Low,
Phys. Rev. {\bf D 19}, 2105 (1979).
\bibitem{[27]} 
N. N. Achasov, S. A. Devyanin and G. N. Shestakov,
Sov. J. Nucl. Phys. {\bf 32}, 566 (1980); Phys. Lett. {\bf B 96}, 168
(1980); Phys. Lett. {\bf B 108}, 134 (1982); Z. Phys. {\bf C 16}, 55 (1982);
N. N. Achasov and G. N. Shestakov, Z. Phys. {\bf C 41}, 309 (1988);
N. N. Achasov and V. V. Gubin, Phys. Rev. {\bf D 56}, 4084 (1997).
\bibitem{[28]}
S. L. Adler,
Phys. Rev. {\bf 177}, 2426 (1969);
J. S. Bell and R. Jackiw,
Nuovo Cim. {\bf A 60}, 47 (1969);
R. Jackiw,
in {\it LECTURES ON CURRENT ALGEBRA AND ITS APPLICATIONS}, 
Princeton University Press, Princeton, New Jersey, 1972;  
R. A. Bertlemann, 
in {\it
ANOMALIES IN QUANTUM FIELD THEORY}, Oxford Science Publications,
Clarendon Press--Oxford, 1996.
\bibitem{[29]}
I. S. Gertsein and R. Jackiw,
Phys. Rev. {\bf 181}, 1955 (1969).
\bibitem{[30]}
S. L. Adler and W. A. Bardeen,
Phys. Rev. {\bf 182}, 1517 (1969).
\bibitem{[31]}
R. W. Brown, C. C. Shih and B. L. Yang,
Phys. Rev. {\bf 186}, 1491 (1969).
\bibitem{[32]} 
R. Jackiw, 
in {\it CURRENT ALGEBRA AND ANOMALIES}, 
S. B. Treiman,
R. Jackiw, B. Zumino and E. Witten (eds), World Scientific, Singapore,
p.81 and p.211; 
N. S. Manton, 
Ann. of Phys. (NY) {\bf 159}, 220 (1985);
N. Ogawa, 
Progr. Theor. Phys. {\bf 90}, 717 (1993); 
R. A. Bertlemann, 
in {\it
ANOMALIES IN QUANTUM FIELD THEORY}, Oxford Science Publications,
Clarendon Press--Oxford, 1996, pp.227--233 and references therein.
\bibitem{[33]}
J. D. Walecka,
Ann. Phys. (NY) {\bf 83}, 121 (1974);
C. J. Horowitz and  B. D. Scrot,
Nucl. Phys. {\bf A 368}, 503 (1981);
Phys. Lett. {\bf B 140}, 181 (1984);
 R. J. Perry,
Phys. Lett. {\bf B 182}, 269 (1986);
 T. D. Cohen,
Phys. Rev. {\bf C 45}, 833 (1992); 
J. C. Caillon and J. Labarsouque,
Phys. Lett. {\bf B 311}, 19 (1993);
 J. Caro, E. Ruiz Arriola and  L. L. Salcedo,
Phys. Lett. {\bf B 383}, 9 (1996);
M Matsuzaki,
Phys. Rev. {\bf C 58}, 3407 (1998).
\bibitem{[34]} 
J. M. Blatt and V. F. Weisskopf, in {\it
THEORETICAL NUCLEAR PHYSICS}, John Wiley $\&$ Sons, New York Chapman
$\&$ Hall Ltd, London, 1952.
\bibitem{[35]} 
A. N. Ivanov, H. Oberhummer, N. I. Troitskaya and M. Faber, 
{\it Neutron--proton radiative capture, photo--magnetic and
anti--neutrino disintegration of the deuteron in the relativistic
field theory model of the deuteron}, nucl--th/9908080, August  1999.
\bibitem{[36]} 
A. N. Ivanov, H. Oberhummer, N. I. Troitskaya and M. Faber, 
{\it Solar proton burning, neutrino disintegration of the
deuteron and pep process in the relativistic field theory model of the
deuteron}, nucl--th/9910021, October 1999.
\bibitem{[37]} 
C. E. Rolfs and W. S. Rodney, 
in {\it CAULDRONS IN THE
COSMOS}, the University of Chicago Press, Chicago and London, 1988.
\bibitem{[38]}
R. J. Oakes,
Phys. Rev. {\bf 131}, 2239 (1963).
\end{thebibliography}
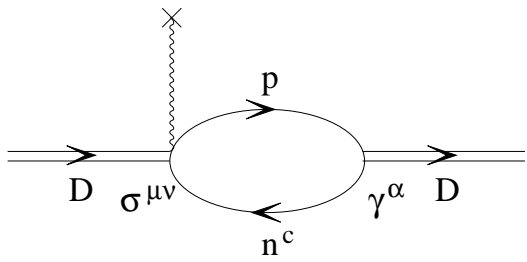
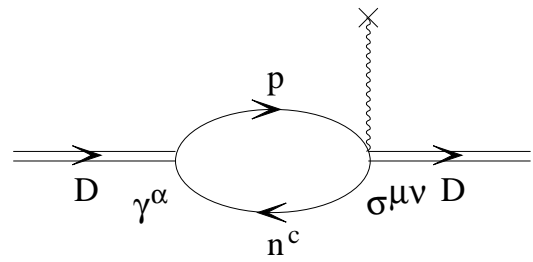
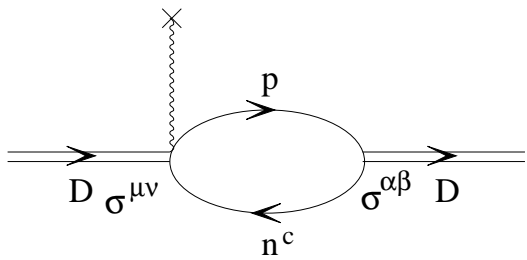
\end{document}